\documentclass[submission, Phys]{SciPost}

\hypersetup{
    colorlinks,
    linkcolor={red!50!black},
    citecolor={blue!50!black},
    urlcolor={blue!80!black}
}

\usepackage[bitstream-charter]{mathdesign}
\usepackage{comment}
\usepackage{booktabs}
\usepackage{array}
\usepackage{cleveref}
\usepackage[most]{tcolorbox}
\newtcolorbox{algobox}[1]{
  enhanced,
  breakable,
  colback=white,
  colframe=black!75,
  colbacktitle=black!75,
  coltitle=white,
  fonttitle=\bfseries,
  title={#1},
  boxrule=0.7pt,
  arc=1.5pt,
  left=7pt,
  right=7pt,
  top=4pt,
  bottom=5pt,
  before skip=8pt,
  after skip=8pt,
}

\newcommand{\Rbb}{R_{b\bar{b}}}
\newcommand{\rc}{r_{\mathrm{c}}}
\newcommand{\dR}{\Delta R}
\newcommand{\dphi}{\Delta\phi}
\newcommand{\pT}{p_{\mathrm{T}}}

\newcommand{\phim}{\phi_{\mathrm{m}}}
\newcommand{\etam}{\eta_{\mathrm{m}}}
\newcommand{\GNN}{\mathrm{GNN}}
\DeclareMathOperator{\atan}{atan}

\DeclareSymbolFont{usualmathcal}{OMS}{cmsy}{m}{n}
\DeclareSymbolFontAlphabet{\mathcal}{usualmathcal}

\begin{document}

\begin{center}{\Large \textbf{
Improving Sensitivity to Rare Processes at the LHC with Geodesic-Aware Graph Neural Networks\\
}}\end{center}

\begin{center}
Benjamin Fuks\textsuperscript{1$\ast$},
Ahmed Hammad\textsuperscript{2$\dagger$},
Adil Jueid\textsuperscript{2$\S$} and 
Mihoko Nojiri\textsuperscript{3,4,5$\Vert$}
\end{center}

\begin{center}
{\bf 1} Laboratoire de Physique Th\'{e}orique et Hautes \'{E}nergies (LPTHE), UMR 7589, Sorbonne Universit\'{e} 
\& CNRS, 4 place Jussieu, 75252 Paris Cedex 05, France.
\\
{\bf 2} Center of AI and Natural science, KIAS, 02455 Seoul, Republic of Korea.
\\
{\bf 3} Theory Center, IPNS, KEK, 1-1 Oho, Tsukuba, Ibaraki 305-0801, Japan.
\\
{\bf 4} The Graduate University of Advanced Studies (Sokendai), 1-1 Oho, Tsukuba, Japan.
\\
{\bf 5} Kavli IPMU (WPI), University of Tokyo, 5-1-5 Kashiwanoha, Chiba 277-8583, Japan.
\\

${}^\ast$ {\small \sf fuks@lpthe.jussieu.fr}
${}^\dagger$ {\small \sf hammad@kias.re.kr}
${}^\S$ {\small \sf adiljueid@kias.re.kr}
${}^\Vert$ {\small \sf nojiri@post.kek.jp} 
\end{center}

\begin{center}
\today
\end{center}


\section*{Abstract}
{\bf
Rare processes at the Large Hadron Collider often lead to final states with quarks, for which the dominant backgrounds contain the same reconstructed objects as the signal. The colour structure of the hard process, imprinted on the pattern of soft QCD radiation, carries information that is largely left unused by standard analyses. We introduce a geometric representation of this colour flow by mapping hadronic constituents onto the unit sphere through a modified inverse stereographic projection centred on a pair of jets selected as a colour-singlet resonance candidate. The jet separation defines a circular region whose interior and exterior are mapped to opposite hemispheres, allowing a graph neural network to exploit the resulting radiation pattern using geodesic distances. As a benchmark, we consider flavour-changing neutral interactions of the top quark with the Higgs boson at $\sqrt{s}=13.6$~TeV, through associated $tH$ production and $t\bar t$ production with one top quark undergoing a $t\to qH$ decay, where $q=u,c$, focusing on subsequent $H\to b\bar b$ and leptonically top quark decays in both channels. Relative to a baseline graph neural network operating in the $(\eta,\phi)$ plane, the geodesic-aware network improves the expected 95\% confidence-level limits on $\mathrm{BR}(t\to uH)$ and $\mathrm{BR}(t\to cH)$ by about an order of magnitude, to $1.1\times10^{-4}$ and $8.3\times10^{-5}$ for an assumed integrated luminosity of $139~\mathrm{fb}^{-1}$. The approach can also be applied to searches for other colour-singlet resonances decaying into quarks or gluons.
}

\vspace{10pt}
\noindent\rule{\textwidth}{1pt}
\tableofcontents\thispagestyle{fancy}
\noindent\rule{\textwidth}{1pt}
\vspace{10pt}

\section{Introduction}
\label{sec:intro}

Since the discovery of the Higgs boson, the physics programme of the Large Hadron Collider (LHC) has placed an increasing emphasis on rare production processes and experimentally challenging decay channels, many of which involve quarks in the final state and yield signals that must be extracted from Standard Model (SM) backgrounds with much larger rates. For instance, the decay of the Higgs boson into bottom quarks, although it has the largest branching fraction, was observed only in 2018~\cite{ATLAS:2018kot, CMS:2018nsn}, and its decay into charm quarks is still being searched for~\cite{CMS:2022psv, ATLAS:2024yzu}. Furthermore, the Higgs-boson self-coupling is probed through Higgs boson pair production in the four-$b$ final state~\cite{ATLAS:2023qzf, CMS:2022cpr}, the top quark Yukawa coupling through $t\bar tH$ production with an $H\to b\bar b$ decay~\cite{ATLAS:2021qou}, while four-top-quark production was first observed in 2023~\cite{ATLAS:2023ajo, CMS:2023ftu}. Beyond these SM measurements, flavour-changing neutral-current (FCNC) interactions of the top quark with a $Z$ boson, a photon, a gluon or a Higgs boson are the subject of dedicated searches~\cite{ATLAS:2023qzr, ATLAS:2022per, CMS:2023bjm, ATLAS:2021amo, ATLAS:2018jqi, CMS:2021ncc, ATLAS:2023mcc, ATLAS:2023ujo, CMS:2024ubt}.

A common difficulty in these analyses is that the small signal yields must be extracted from backgrounds that can contain the same reconstructed objects, including several jets and often multiple $b$-tagged jets. For example, $t\bar tH$ production with $H\to b\bar b$ competes with $t\bar t\, b\bar b$ production, Higgs-boson pair production in the four-$b$ final state with QCD multijet production, and associated $VH$ production with $H\to c\bar c$ with vector-boson production in association with jets. Although invariant masses, angular separations and other kinematic observables provide substantial discrimination, the overlap between the signal and the background distributions motivates the use of additional information beyond the reconstructed event kinematics. Such information could originate from the colour structure of the hard process, which influences the pattern of soft QCD radiation through interference between emissions from different partons~\cite{Maltoni:2002mq, Gallicchio:2010sw}. In particular, the quark-antiquark pair produced in the decay of a colour-singlet resonance, such as a Higgs or a $Z$ boson, forms a colour dipole whose radiation pattern differs from that of a pair produced through gluon splitting or other QCD mechanisms, where colour connections to the rest of the event modify the distribution of radiation between and around the jets. Observable consequences of these colour connections persist after hadronisation, as demonstrated by measurements in top-quark pair events at the Tevatron and the LHC~\cite{D0:2011lzz, ATLAS:2015ytt, ATLAS:2018olo}, and can be exploited through observables such as jet pull~\cite{Gallicchio:2010sw}, jet dipolarity~\cite{Hook:2011cq} and the colour ring~\cite{Buckley:2020kdp}, or through image-based learning of radiation patterns~\cite{Kim:2019wns}. By probing correlations in the radiation pattern that are not fully specified by the reconstructed hard kinematics, colour-flow observables can therefore provide complementary discrimination against backgrounds with similar reconstructed final states.

Extracting this information from the many particles in an event is a natural application of machine learning, whose role in collider event classification has expanded considerably in recent years~\cite{Larkoski:2017jix, Albertsson:2018maf, Radovic:2018dip, Guest:2018yhq}. Successful approaches include particle flow networks~\cite{Komiske:2018cqr}, classifiers based on energy flow polynomials~\cite{Komiske:2017aww}, graph neural networks (GNNs)~\cite{Shlomi:2020gdn, Qu:2019gqs}, transformers~\cite{Qu:2022mxj,Esmail:2025kii} and Lorentz-equivariant networks~\cite{Gong:2022lye}, which exploit different representations of particle-level information for jet tagging and event classification. These developments illustrate the value of incorporating the symmetries of particle-level data into the learning architecture, following the broader principles of geometric deep learning~\cite{Bronstein:2021mdi}. This geometric perspective also extends to the representation of collider events themselves, through approaches based on optimal-transport distances~\cite{Komiske:2019fks, Cai:2020vzx, Komiske:2020qhg, Cai:2021hnn} and on Riemannian metrics defined on relativistic phase space~\cite{Cai:2024xnt}. More directly related to colour flow, spherical representations based on inverse stereographic projection have been introduced to emphasise the radiation patterns associated with different colour structures~\cite{Hammad:2022msl}, while the joint use of radiation patterns and event kinematics has been investigated in multimodal networks~\cite{Ban:2023jfo}. 

We build on this geometric treatment of colour flow by combining a modified inverse stereographic projection (ISP) of individual hadronic constituents onto the unit sphere $S^2$ with a geodesic-aware GNN, using FCNC interactions of the top quark with the Higgs boson as a benchmark. We consider associated $tH$ production and $t\bar t$ production with one top quark undergoing the FCNC decay $t\to qH$, where $q=u,c$. In both cases, the decay $H\to b\bar b$ provides the colour-singlet quark pair whose radiation pattern motivates the mapping. For each event, the mapping is centred on the midpoint of the two $b$-tagged jets selected as the Higgs-boson candidate, with their separation setting the diameter of a circular region in the $(\eta,\phi)$ plane. Hadronic constituents inside and outside this region are mapped by construction to the southern and northern hemispheres of the unit sphere, respectively, so that differences in the distribution of radiation relative to the candidate become hemispheric asymmetries. A $k$-nearest-neighbour graph is then constructed using the intrinsic geodesic distance on $S^2$ both to identify neighbouring constituents and to define the edge attributes.

The choice of this benchmark is further motivated by the strong suppression of top-quark FCNC decays in the SM, where the $t\to qH$ decay process is forbidden at tree level and suppressed at one loop by the Glashow-Iliopoulos-Maiani mechanism, yielding branching fractions of order $10^{-17}$ and $10^{-15}$ for the up- and charm-quark channels, respectively~\cite{Eilam:1990zc, Mele:1998ag, Aguilar-Saavedra:2004mfd, Bardhan:2016txk}. More generally, rates for top-quark FCNC processes can be enhanced by many orders of magnitude in extensions of the SM, including two-Higgs-doublet models~\cite{Atwood:1996vj, Botella:2015hoa, Abbas:2015cua, Baum:2008qm, Arroyo-Urena:2019qhl} and three-Higgs-doublet models~\cite{Coleppa:2026fdj}, supersymmetric models~\cite{Dedes:2014asa, Cao:2007dk, Lopez:1997xv, Eilam:2001dh}, extra-dimensional setups~\cite{Gao:2013fxa, Diaz-Furlong:2016ril, Dey:2016cve}, left-right symmetric constructions including supersymmetric realisations~\cite{Gaitan:2004by, Frank:2005vd, Frank:2023fkc}, models with scalar leptoquarks~\cite{Bolanos:2019dso}, additional dark-sector particles~\cite{Liu:2021crr, Jueid:2024cge} or vector-like quarks~\cite{Aguilar-Saavedra:2002lwv, Balaji:2021lpr, Crivellin:2022fdf}, as well as composite-Higgs scenarios~\cite{Azatov:2014lha}. An observation of these decays at experimentally accessible rates would therefore provide a clear indication of new physics. 

For the Higgs channel, LHC Run-2 combinations of several decay channels by ATLAS and CMS yield expected 95\% confidence-level upper limits on the branching ratio $\mathrm{BR}(t\to qH)$ in the range of $(2-4) \times 10^{-4}$~\cite{ATLAS:2023ujo, CMS:2024ubt}. In the present study, which extends our earlier deep-learning analysis of top-quark FCNC couplings to the photon~\cite{Fuks:2025qgh}, we select events in which the accompanying top quark decays leptonically, with the dominant background arising from $t\bar t$ production in association with jets. To assess the discrimination achievable in this setting, we compare two multimodal classifiers that combine a particle-level GNN with a multilayer perceptron (MLP) processing high-level kinematic observables: a geodesic-aware GNN (GA-GNN) using the spherical representation, and a baseline `plain' GNN constructed in the $(\eta,\phi)$ plane with $\Delta R$ edge attributes. Both classifiers use the same architecture for the kinematic branch and classification head, while their GNN branches use edge-conditioned convolutions (NNConv)~\cite{SimonovskyK17, Gilmer:2017mpnn} with the same hidden-layer configuration, optimiser and learning-rate schedule. We translate the classifier outputs into expected exclusion limits using three complementary statistical treatments: an asymptotic estimate, an exact Poisson calculation for low event counts and a profile-likelihood analysis incorporating the specified systematic uncertainties. For simulated proton-proton collisions at $\sqrt{s}=13.6$~TeV and an assumed integrated luminosity of $139~\mathrm{fb}^{-1}$, the GA-GNN yields expected limits on $\mathrm{BR}(t\to uH)$ and $\mathrm{BR}(t\to cH)$ that are found about an order of magnitude stronger than those obtained with the plain GNN.

The remainder of this paper is organised as follows. In section~\ref{sec:colourflow} we review the physics of colour flow and introduce the spherical representation. Section~\ref{sec:fcnc} describes the new physics benchmark scenario, the simulation of signal and background events, and the event selection and reconstruction. The two classifiers are presented in section~\ref{sec:ml}, and the statistical frameworks and expected limits are given in section~\ref{sec:results}. We summarise our findings and discuss future directions in section~\ref{sec:conclusion}.

\section{Encoding colour flow on the unit sphere}
\label{sec:colourflow}

\subsection{Colour flow in QCD}
\label{subsec:cf_physics}

The pattern of soft QCD radiation in an event is governed by the colour connections between the coloured partons of the hard process. Since QCD conserves the $SU(3)_c$ colour symmetry, these connections are fixed by the way the colour indices flow through the hard scattering amplitude, and they can be made explicit in the colour flow representation of the QCD amplitudes~\cite{Maltoni:2002mq, Hagiwara:2010vk, Kilian:2012pz}. In the soft limit, the angular distribution of a soft gluon of four-momentum $k$ emitted coherently by a pair of colour connected partons $i$ and $j$ of respective four-momenta $p_i$ and $p_j$ is controlled by the eikonal (antenna) function
\begin{equation}
  W_{ij}(k) = \frac{p_i \cdot p_j}{(p_i \cdot k)(p_j \cdot k)}\,.
  \label{eq:antenna}
\end{equation}
It is large when $k$ is collinear to $p_i$ or to $p_j$, and it remains sizeable in the angular region between the two partons: a colour connected pair radiates as a dipole into the region delimited by the directions of its two partons. By contrast, for partons that are not directly colour connected, radiation in the region between them depends on their connections to the other partons and need not exhibit the same inter-parton enhancement, and soft radiation therefore follows the colour connections of the hard process. This pattern survives hadronisation and can be observed in the angular distribution of low momentum  hadrons~\cite{Gallicchio:2010sw, Kim:2019wns}, and it has been measured in top quark pair events~\cite{D0:2011lzz, ATLAS:2015ytt, ATLAS:2018olo}.

To illustrate how these colour-flow differences can be exploited, we consider a top-quark FCNC benchmark scenario in which the signal comprises associated $tH$ production and $t\bar t$ production with one top quark undergoing an FCNC decay $t\to qH$, where $q=u,c$. In both channels, we consider the subsequent decay $H\to b\bar b$, together with a leptonic decay of the accompanying top quark. Although these signals and their backgrounds can share similar hard kinematics with a charged lepton, missing transverse momentum and several $b$-jets, the colour connections of the jets forming the Higgs candidate can lead to different patterns of soft radiation.

In the signal, the Higgs boson is a colour singlet. In the decay $H\to b\bar b$, the $b$ and $\bar b$ quarks carry conjugate $SU(3)_c$ charges and are connected by a single colour string~\cite{Andersson:1983ia}. With a string tension of about $1~\mathrm{GeV/fm}$, the string fragments along its length and produces soft hadrons between the two $b$-quarks. This colour connection enhances radiation in the inter-parton region, which we probe using a cone of radius $\rc=\Delta R_{b\bar b}/2$ centred on the midpoint of the $b\bar b$ pair in the $(\eta,\phi)$ plane. The top quark decays semileptonically, $t\to Wb\to\ell\nu b$, and contributes a further $b$-jet, a charged lepton and missing transverse momentum. Here, the $b$ quark from the top decay inherits the colour connections of its parent, linking it to the rest of the event rather than to the colour-singlet Higgs decay system.

In the dominant background, $t\bar t$ production in association with jets, the two $b$-tagged jets forming the Higgs candidate can originate from gluon splitting, $g\to b\bar b$, or from different parts of the event. In either case they are colour-connected to the rest of the event rather than exclusively to each other, and soft radiation is deposited both inside and outside the $b\bar b$ cone, without any colour-singlet dipole structure. Other backgrounds containing a genuine colour-singlet $b\bar b$ pair require separate consideration. In $t\bar tH$ production with $H\to b\bar b$, and in $t\bar tZ$ production with $Z\to b\bar b$, a genuine colour-singlet $b\bar b$ pair is present, but the additional colour connections of the $t\bar t$ pair fill the event with radiation which can dilute the characteristic inside-cone pattern and make the correct identification of the pair more difficult. Finally, the $t\bar tW$, $tW$ and $tj$ processes contain instead no colour singlet resonance decaying into $b\bar b$.

\subsection{Event-centred frame and radiation asymmetry}
\label{subsec:centering}

To characterise the radiation around the Higgs candidate, we describe each event in a common frame centred on the $b\bar b$ pair. We take the two $b$-jets, denoted $b_1$ and $b_2$, and define their midpoint in the $(\eta,\phi)$ plane, 
\begin{align}
  \eta_m &= \frac{\eta_{b_1} + \eta_{b_2}}{2}\,,
  \label{eq:eta_m}\\[4pt]
  \phi_m &= \phi_{b_1} \oplus \frac{\dphi(\phi_{b_2},\phi_{b_1})}{2}\,,
  \label{eq:phi_m}
\end{align}
where $\oplus$ denotes addition modulo $2\pi$ and
$\dphi(\phi_1,\phi_2) = [(\phi_1-\phi_2+\pi)\bmod 2\pi]-\pi$ is the signed
azimuthal difference. All hadrons of the event are then expressed in the
centred coordinates
\begin{equation}
  \eta_c = \eta - \eta_m\,,
  \qquad
  \phi_c = \dphi(\phi,\phim)\,,
  \label{eq:centering}
\end{equation}
and the reference radius is set to half the separation of the pair in the $(\eta,\phi)$ plane,
\begin{equation}
  \rc = \frac{\Delta R_{b\bar{b}}}{2}
      = \frac{1}{2}\sqrt{(\Delta \eta_{b_1 b_2})^2 + (\Delta \phi_{b_1 b_2})^2}\,.
  \label{eq:rc}
\end{equation}
In the centred frame, the two $b$-jets lie symmetrically about the origin, at a distance $\rc$ from it. The disc of radius $\rc$ contains the region between them, that we populate with both charged and neutral hadrons, while the isolated leptons and prompt photons are excluded. Centring removes the absolute pseudorapidity and azimuthal position of the Higgs candidate from these coordinates, allowing radiation patterns from different events to be compared directly in a common coordinate system.

\begin{figure}
    \centering
    \includegraphics[width=0.95\linewidth]{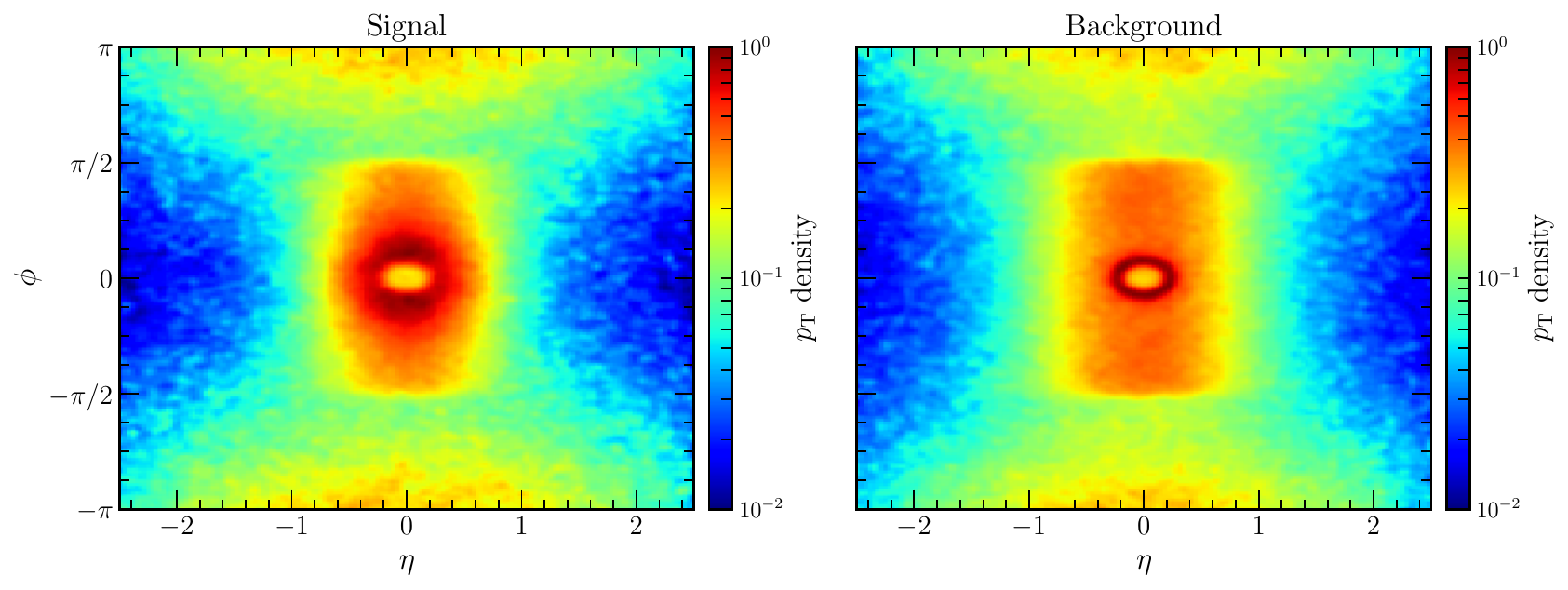}
    \caption{Accumulated average transverse momentum density in the centred $(\eta_c,\phi_c)$ plane for 50\,000 $tH$ signal events (left) and 50\,000 background events (right). The background is composed of the processes of \cref{tab:xs}, combined using the corresponding per-event weights. The density is constructed after the event selection described in \cref{sec:event_selection}, and each density is normalised to its maximum.}
    \label{fig:cumulativeAv}
\end{figure}

A simple observable that captures the expected difference is the inside/outside asymmetry of the radiation with respect to the $b\bar b$ cone. With the radial distance
\begin{equation}
  r = \sqrt{\eta_c^2 + \phi_c^2}
  \label{eq:radius}
\end{equation}
of a hadron in the centred frame, we define
\begin{equation}
  \mathcal{A}_{\rm cf}
  = \frac{\displaystyle\sum_{r_i < \rc} \pT^i
          - \displaystyle\sum_{r_i > \rc} \pT^i}
         {\displaystyle\sum_{i} \pT^i}\,,
  \label{eq:asymmetry}
\end{equation}
which is positive when the scalar transverse-momentum sum inside the cone exceeds that outside it, and negative in the opposite case. \Cref{fig:cumulativeAv} shows the accumulated average transverse momentum density of the hadrons in the centred frame for signal and background events, which exhibit different average spatial profiles around the reconstructed Higgs candidate: signal events display a filled cone profile, qualitatively consistent with radiation associated with the colour-singlet Higgs decay, whereas background events show a more uniform and extended pattern, which reflects the different colour connections in the relevant processes.

A single number such as $\mathcal{A}_{\rm cf}$, however, discards most of the two dimensional pattern of the radiation. The approach developed in this work instead represents the hadronic constituents as a point cloud on the unit sphere, allowing a graph neural network to exploit their angular distribution.

\subsection{Modified inverse stereographic projection}
\label{subsec:isp}

\begin{figure}[!t]
    \includegraphics[width=1.\linewidth]{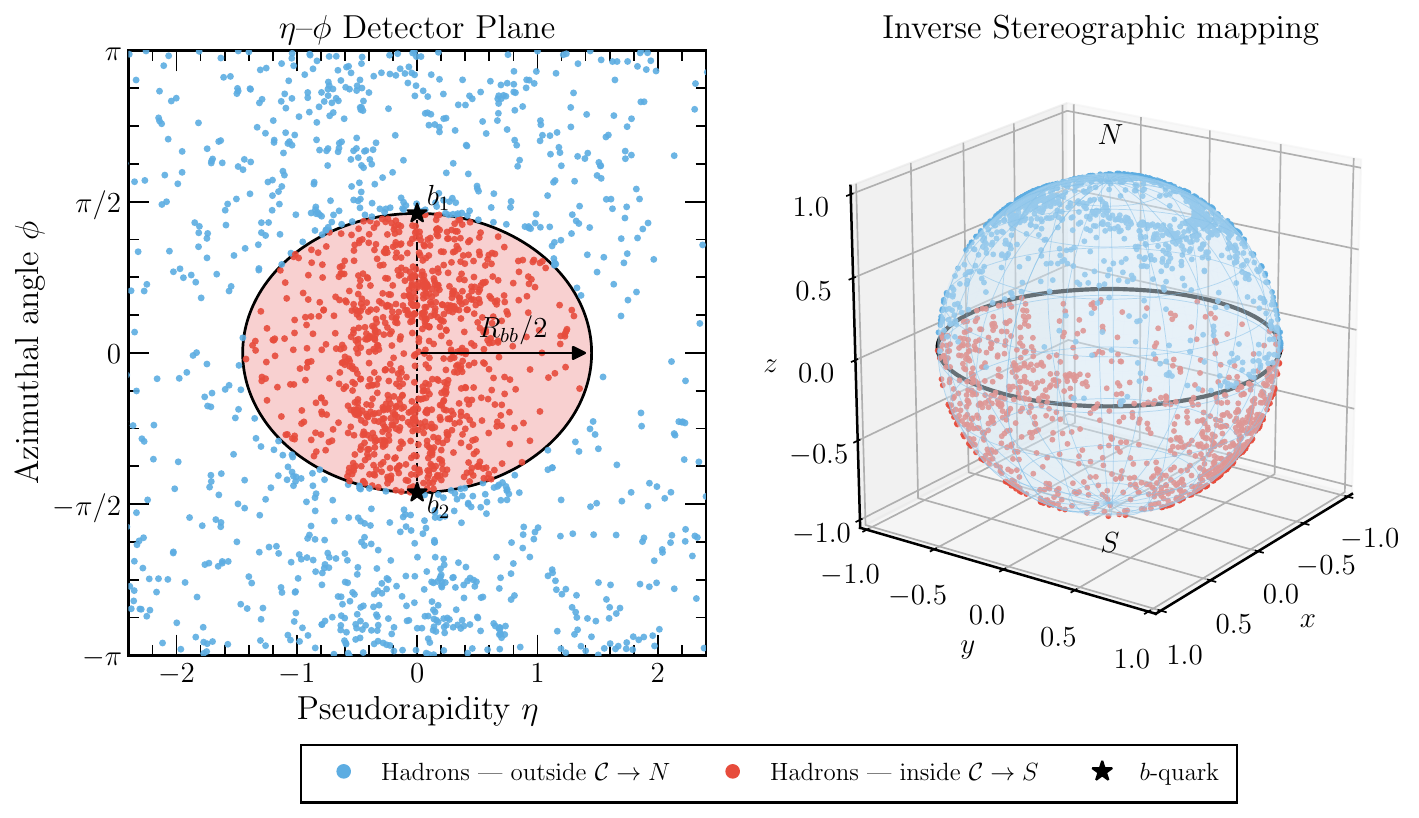}
    \caption{Schematic illustration of the modified inverse stereographic mapping from the centred $(\eta_c,\phi_c)$ detector plane (left) to the unit sphere (right). Hadrons inside (outside) the Higgs cone $\mathcal{C}$ are mapped to the southern (northern) hemisphere, while the two Higgs $b$-jets lie on the equator.}
    \label{fig:projection}
\end{figure}

Given the centred coordinates $(x,y) = (\eta_c,\phi_c)$ of a hadron and its radial distance $r=\sqrt{x^2+y^2}$ from \cref{eq:radius}, we map it to a point in $\mathbb{R}^3$,
\begin{equation}
  X = \frac{2x}{r^2+1}\,,
  \qquad
  Y = \frac{2y}{r^2+1}\,,
  \qquad
  Z = \tanh(r - \rc)\,.
  \label{eq:isp_xyz}
\end{equation}
By contrast, the classical inverse stereographic projection maps $(x,y)\in\mathbb{R}^2$ onto the sphere without its north pole, with $Z = (r^2-1)/(r^2+1)$. Here, $Z$ is replaced by $\tanh(r-\rc)$ so that, after normalisation, hadrons inside the $b\bar b$ cone ($r<\rc$) are mapped to the southern hemisphere, while those outside it ($r>\rc$) are mapped to the northern hemisphere and the cone boundary ($r=\rc$) lies on the equator. This geometric separation is motivated by the radiation pattern associated with the colour-singlet Higgs decay, which favours the region between its two decay products, and by contributions from the top-quark system, initial-state radiation and the underlying event that can extend beyond this region. The spherical coordinates $(\theta,\phi_s)$ are extracted after normalising the vector $(X,Y,Z)$ defined by the modified mapping to unit length,
\begin{equation}
  \hat{n} = (\hat{n}_X,\hat{n}_Y, \hat{n}_Z) = \frac{(X,Y,Z)}{\|(X,Y,Z)\|}\,,
  \qquad
  \theta = \arccos(\hat{n}_Z)\,,
  \qquad
  \phi_s = \atan2(\hat{n}_Y,\hat{n}_X)\,,
  \label{eq:sphere_angles}
\end{equation}
which places every hadron exactly on $S^2$, regardless of the magnitude of the intermediate vector $(X,Y,Z)$. The mapping is illustrated schematically in \cref{fig:projection}.

\begin{figure}
  \centering
  \includegraphics[width=0.85\linewidth]{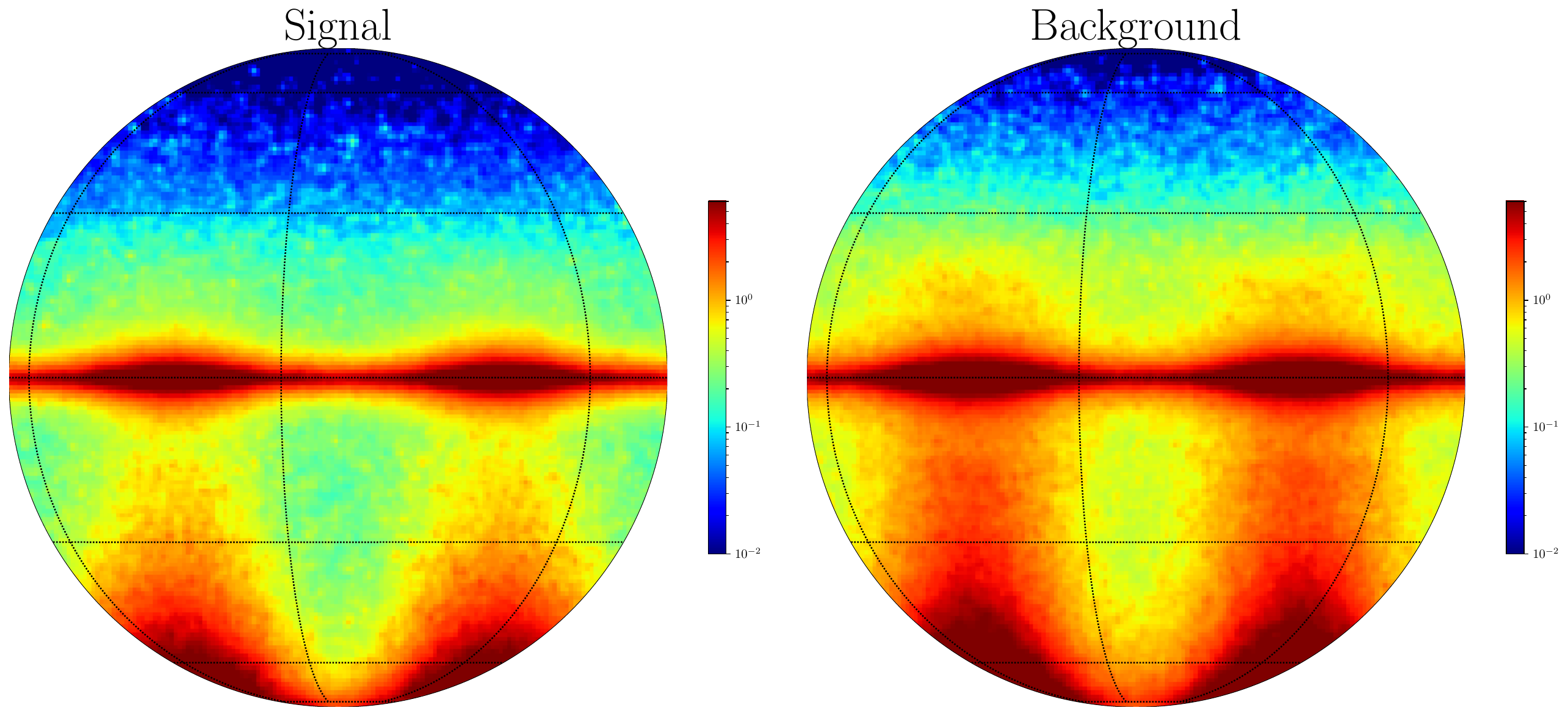}
  \caption{Accumulated average transverse-momentum density on the unit sphere, for 50\,000 $tH$ signal events (left) and 50\,000 background events (right), with the same composition as in \cref{fig:cumulativeAv}. The colour scale gives the transverse-momentum density on a logarithmic scale, and the equatorial enhancement thus traces hadrons near the boundary of the reconstructed Higgs cone.}
  \label{fig:mapping}
\end{figure}

To better highlight how the projection represents the radiation patterns, we accumulate the $\pT$-weighted density of hadrons on the sphere over all events of each class and project it onto the $(\theta,\phi_s)$ plane in \cref{fig:mapping}. For the signal, the soft hadrons from the fragmentation of the colour singlet string populate predominantly the southern hemisphere. For the dominant contributions to the background, colour connections to the rest of the event can lead to a more uniform distribution, with a substantial northern component from radiation outside the cone produced by the colour charged top quarks, by initial-state gluons and by the underlying event. This north/south asymmetry therefore provides a geometric representation of the differences bwtween the signal and the background radiation patterns, and is the primary feature that the GA-GNN is designed to exploit in \cref{sec:ml}.

\subsection{Coordinate conventions and geometric properties}
\label{subsec:invariance}

The spherical representation defined by eqs.~\eqref{eq:centering}--\eqref{eq:sphere_angles} has three properties that are desirable for a colour flow observable:
\begin{enumerate}
  \item \textbf{Longitudinal boosts.}
        Differences in rapidity are invariant under boosts along the beam axis, and so are differences in pseudorapidity for massless particles. Subtracting $\etam$ in eq.~\eqref{eq:centering} therefore makes the centred coordinates invariant under a common longitudinal boost in the massless approximation.
  \item \textbf{Azimuthal rotations.}
        Subtracting $\phi_m$ centres each event at the midpoint of the reconstructed $b\bar b$ pair in azimuth,  removing the dependence on its absolute azimuthal position in the detector.
  \item \textbf{Size of the dipole.}
        A hadron at $r = \alpha\rc$ is mapped to $Z = \tanh\big((\alpha-1)\rc\big)$, whose sign depends only on the fractional distance $\alpha = r/\rc$ and not on the absolute size $\rc$ of the cone. Two signal events with different Higgs-boson transverse momenta, and hence different opening angles $\Rbb$, therefore place hadrons at the same fractional distance from the $b\bar b$ axis in the same hemisphere, even though the precise polar angle still depends on $\rc$.
\end{enumerate}
Together, these properties express the event geometry relative to the Higgs candidate and make the hemisphere assignment depend only on the fractional distance $r/\rc$, while the detailed spherical coordinates retain a dependence on $\rc$.

\section{Benchmark: top quark FCNC coupling to the Higgs boson}
\label{sec:fcnc}

We now describe in detail the top FCNC benchmark scenario introduced above, starting with the relevant effective interaction Lagrangian and the associated signal and background processes in section~\ref{sec:lagrangian}, before presenting the event simulation, selection and reconstruction in section~\ref{sec:event_selection}.

\subsection{Effective Lagrangian, signal processes and backgrounds}
\label{sec:lagrangian}

We parametrise the top quark FCNC coupling to the Higgs boson through the effective interaction Lagrangian \cite{Aguilar-Saavedra:2004mfd, Aguilar-Saavedra:2008nuh, Durieux:2014xla}
\begin{equation}
  \mathcal{L}_{tqH} = - \frac{g}{2\sqrt{2}} \sum_{q=u,c} \bar{q}\,\Big(g^{v}_{qt} + g^{a}_{qt}\,\gamma_5\Big)\, t\, H + \text{H.c.}\,,
  \label{eq:lagrangian}
\end{equation}
where $H$ denotes the physical Higgs boson, $g$ is the $SU(2)_L$ gauge coupling, and $g^{v}_{qt}$ and $g^{a}_{qt}$ are the scalar and pseudoscalar couplings, which in a gauge-invariant formulation arise from dimension-six operators after electroweak symmetry breaking~\cite{Buchmuller:1985jz, Grzadkowski:2010es, Brivio:2017vri}. Neglecting the light-quark masses, this Lagrangian induces the partial decay width
\begin{equation}
  \Gamma(t\to qH) = \frac{\alpha_{\rm EW}}{32 \, s_W^2}\,\Big(|g^{v}_{qt}|^2 + |g^{a}_{qt}|^2\Big)\, m_t
  \left(1-\frac{m_H^2}{m_t^2}\right)^{2},
  \label{eq:width}
\end{equation}
where $\alpha_{\rm EW} = e^2/(4\pi)$ is the fine structure constant, $s_W \equiv \sin\theta_W$ is the sine of the electroweak mixing angle and $m_H$ is the Higgs boson mass. We can compute the corresponding branching ratio as $\mathrm{BR}(t\to qH)\simeq\Gamma(t\to qH)/\Gamma_t$, where $\Gamma_t = 1.32$ GeV is the total top quark width calculated at next-to-next-to-leading order (NNLO) in QCD \cite{Gao:2012ja}. 

\begin{figure}
  \centering \includegraphics[width=0.75\linewidth]{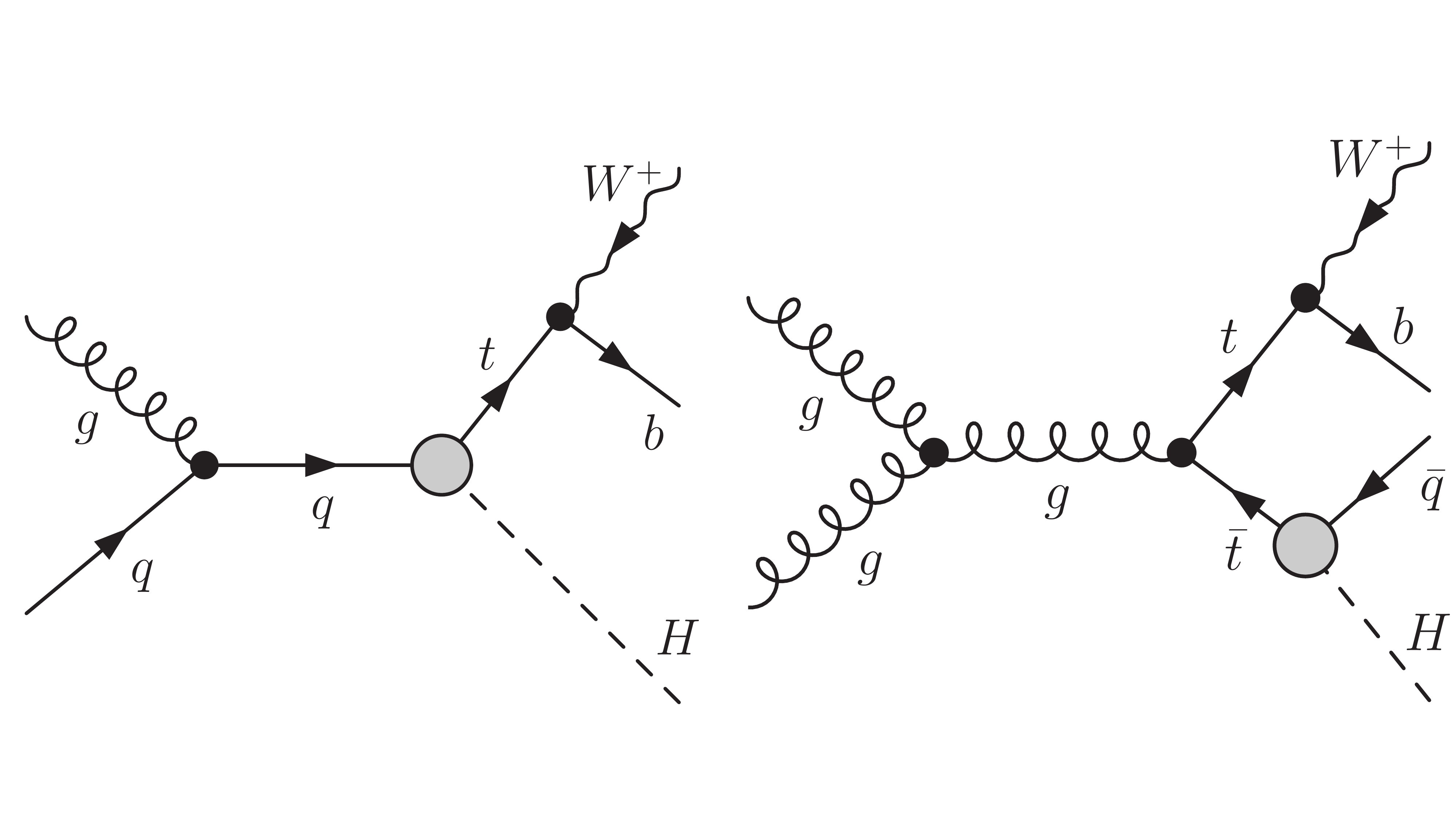} \vspace{-0.4cm}
  \caption{Representative tree-level Feynman diagrams for the FCNC production of 
  a single SM-decaying top quark in association with a Higgs boson (left), and for the 
  SM production of a top-antitop pair where one of the top quarks decays in the SM way 
  and the other through an FCNC process (right). \label{fig:FD:tqH}}
\end{figure}

\begin{figure}[!t]
  \centering
  \includegraphics[width=0.485\linewidth]{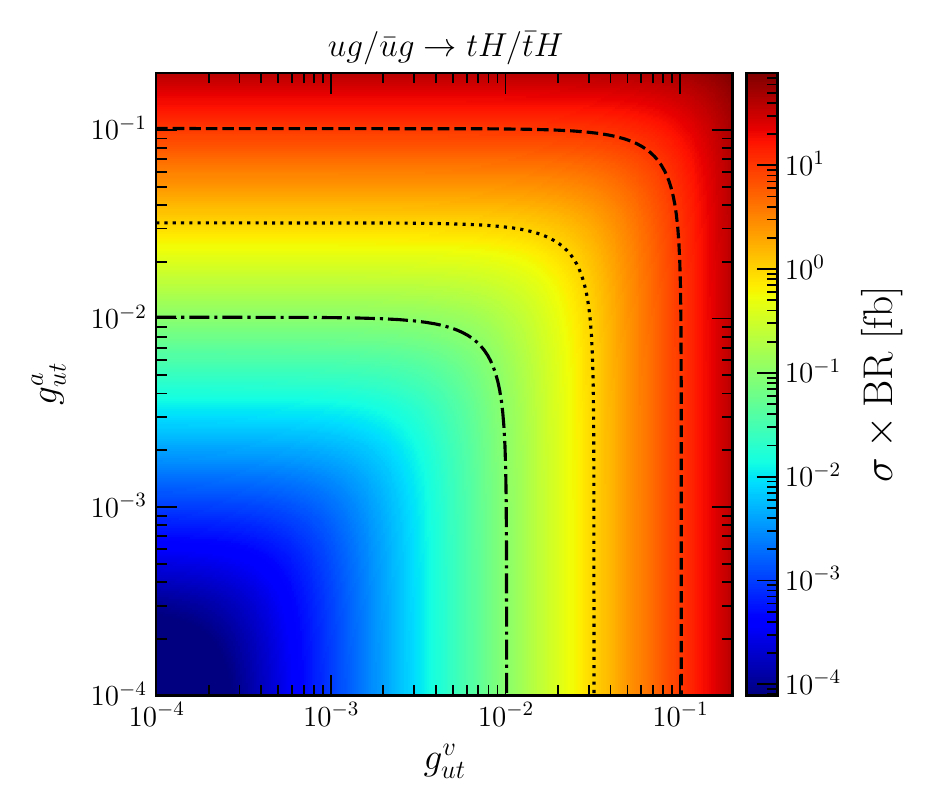}
  \includegraphics[width=0.485\linewidth]{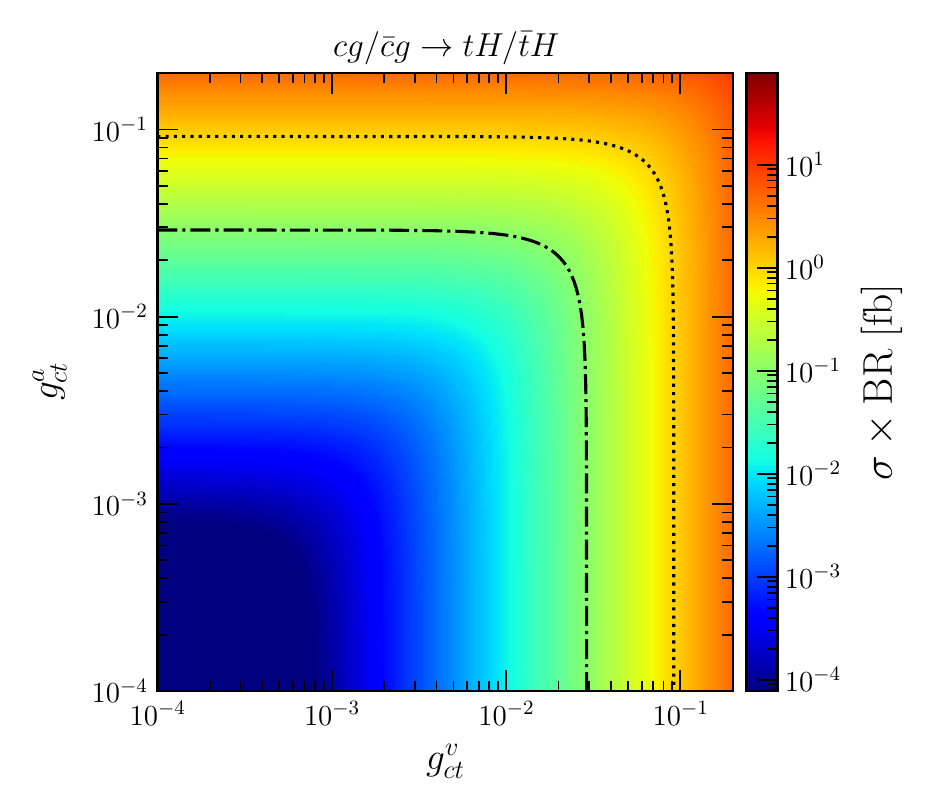}\\
  \includegraphics[width=0.485\linewidth]{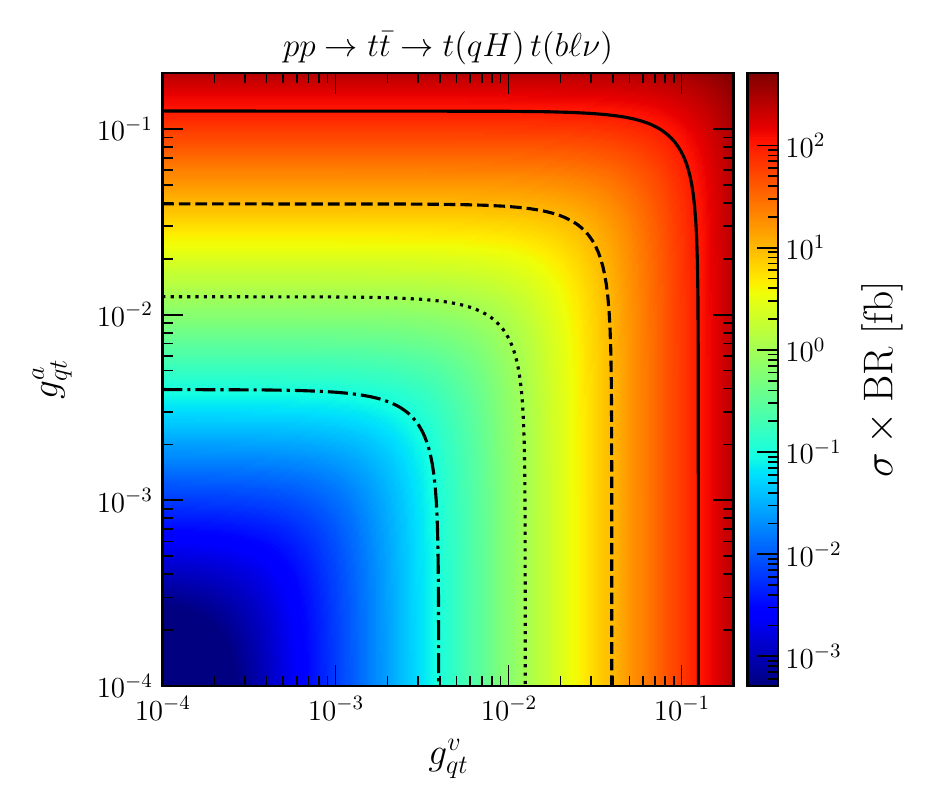}
  \caption{Signal rates including the relevant decay branching fractions, $\sigma\times\mathrm{BR}$, as a function of the scalar and pseudoscalar FCNC couplings $(g_{qt}^{v},g_{qt}^{a})$ at $\sqrt{s}=13.6\,\mathrm{TeV}$. The upper panels show the LO rate for associated $tH$ production for $q=u$ (left) and $q=c$ (right), including the charge-conjugate processes, while the lower panel shows $t\bar t$ production with one FCNC decay $t\to qH$, normalised using the NNLO+NNLL $t\bar t$ cross section. The logarithmic colour scale gives the rate in femtobarns, with black contours at $10^{2}\,\mathrm{fb}$ (solid), $10^{1}\,\mathrm{fb}$ (dashed), $10^{0}\,\mathrm{fb}$ (dotted) and $10^{-1}\,\mathrm{fb}$ (dash-dotted).
  \label{fig:xs}}
\end{figure}

The interaction of eq.~\eqref{eq:lagrangian} also gives rise to the two considered signal processes (see  \cref{fig:FD:tqH} for representative Feynman diagrams). The first involves the production of a single top quark in association with a Higgs boson, $qg\to tH$, in which the initial-state light quark couples to the top quark through the FCNC vertex. The second involves the production of a top-antitop pair, followed by the FCNC decay of one of the top quarks, $pp\to t\bar t$ with $t\to qH$ or $\bar t\to \bar qH$. The corresponding rates are shown in \cref{fig:xs} as a function of the scalar and pseudoscalar couplings $(g_{qt}^{v},g_{qt}^{a})$ at a centre-of-mass energy of $\sqrt{s}=13.6\,\mathrm{TeV}$. For associated production, the partonic cross sections are computed at leading order (LO) in QCD with \textsf{MadGraph5\_aMC@NLO}~\cite{Alwall:2014hca} version 3.4.2 and convoluted with the LO set of \textsf{NNPDF40} parton densities~\cite{NNPDF:2021njg}, accessed through \textsf{LHAPDF} version 6.5.4~\cite{Buckley:2014ana}. The top panels show the resulting rates including the decay branching fractions, thus defined as 
\begin{equation}
\sigma \times {\rm BR} \equiv \Big[\sigma(q g \to t H) + \sigma(\bar{q} g \to \bar{t} H)\Big]
 \times \mathrm{BR}(t\to b\ell\nu)\times\mathrm{BR}(H\to b\bar{b})\,.
\end{equation}
The bottom panel shows the corresponding pair-production rate, obtained from the $t\bar t$ cross section at NNLO in QCD including the resummation of the threshold logarithms at next-to-next-to-leading-logarithmic accuracy (NNLO+NNLL), as computed with \textsf{Top++}~\cite{Czakon:2011xx, Czakon:2013goa} and the NNLO set of parton densities,
\begin{equation}
  \sigma \times {\rm BR} \equiv 2 \times \sigma(p p \to t \bar{t})
   \times \mathrm{BR}(t\to b\ell\nu)\times \mathrm{BR}(t\to qH)\times\mathrm{BR}(H\to b\bar{b})\,.
\end{equation}
For real couplings, neglecting the light-quark masses and assuming that the FCNC decays modify the total width $\Gamma_t$ only marginally, both inclusive rates scale quadratically as $(g_{qt}^{v})^{2} + (g_{qt}^{a})^{2}$, so that the contours of constant rate are circles in a linear $(g^{v},g^{a})$ plane. Moreover, for equal FCNC couplings, the associated production rate is larger for $q=u$ than for $q=c$, reflecting the valence up-quark content of the proton, while the pair-production rate is identical for the two flavours.

For both production modes, we focus on the dominant decay $H\to b\bar b$ and on the semileptonic decay $t\to b\ell\nu$ ($\ell=e,\mu$) of the SM-decaying top quark, and therefore select, in our analysis, events with exactly one isolated charged lepton, missing transverse energy $E_T^{\rm miss}$ and at least three jets identified as originating from $b$-quarks. On the other hand, the $t\bar t$ process yields an additional light flavour jet from the top FCNC decay and QCD radiation populates higher jet multiplicities, so we do not impose an upper bound on the number of jets in the analysis. As already mentioned, a distinctive feature of both signal processes is their colour structure: the Higgs boson is a colour singlet, so its decay products form a colour connected $b\bar b$ pair that is isolated from the rest of the event, whereas in the backgrounds the $b$-quarks are generally colour connected to the top quarks or to the beam remnants~\cite{Maltoni:2002mq,Gallicchio:2010sw}.

The dominant background corresponds to top quark pair production in association with jets, which yields the same lepton and $E_T^{\rm miss}$ signature when one top quark decays leptonically. It enters the signal region of our analysis through events with an additional $b\bar b$ pair from gluon splitting, and through events with charm or light flavour jets misidentified as $b$-jets. The associated production of a top pair with a boson constitutes a smaller but particularly signal-like background: $t\bar t H$ with $H\to b\bar b$ and $t\bar t Z$ with $Z\to b\bar b$ contain a resonant, colour singlet $b\bar b$ pair, as the signal does, while $t\bar t W^\pm$ contributes through its leptonic decays. Single top quark production in the $t$-channel ($pp\to tj$) and in association with a $W$ boson ($pp\to tW^-$), together with their charge conjugate processes, completes the set of simulated backgrounds. We hence do not include  contributions from $W$+jets and QCD multijet production as they are  strongly suppressed by the requirement of three $b$-tagged jets and of an isolated lepton.

The $t\bar t$+jets background is simulated at LO in QCD and events are generated by merging samples with zero, one and two additional matrix-element partons according to the MLM scheme~\cite{Mangano:2006rw, Alwall:2007fs} with a merging scale fixed to 15~GeV. The $t\bar tX$ ($X=H,W,Z$) and single top processes are also computed at LO in QCD, and in all cases we use the five flavour number scheme and dynamical renormalisation and factorisation scales set on an event-by-event basis to
\begin{equation}
\mu_{R, 0} = \mu_{F, 0} = \frac{1}{2} \sum_i \sqrt{p_{T,i}^2 + m_i^2},
\end{equation}
where $p_{T,i}$ is the transverse momentum of the $i$-th particle, $m_i$ is its mass and the sum runs over all final-state partons. These LO calculations use the \textsf{NNPDF40} parton density specified above, and the cross sections of all processes at $\sqrt{s}=13.6$~TeV are listed in table~\ref{tab:xs}. Because the backgrounds differ in cross section by up to three orders of magnitude, the number of generated events $N_{\rm events}$ is chosen per process, and the per-event weight $\omega = \sigma_{\rm LO}/N_{\rm events}$ gives the number of expected events represented by each simulated event, per unit of integrated luminosity. The largest weights occur for the single top $t$-channel and $t\bar t$+jets samples, for which a single simulated event corresponds to about $0.18$ and $0.10$ expected events at $\mathcal{L}=139~\mathrm{fb}^{-1}$, and to about 4.0 and 2.0 expected events at $\mathcal{L}=3~\text{ab}^{-1}$. These two samples therefore dominate the Monte Carlo (MC) statistical uncertainty at tight working points, whereas it is negligible for the signal and for the $t\bar t V$ samples.

\begin{table}
  \setlength\tabcolsep{11pt}\renewcommand{\arraystretch}{1.2}
  \begin{center}
  \begin{tabular}{l ccc}
  Process & $\sigma_{\rm LO}$~[fb] & $N_{\rm events}$ & $\omega = \sigma_{\rm LO} / N_{\rm events}$ [fb] \\
  \toprule
  $p p \to t ~\bar{t}~H$  & $4.92 \times 10^1$ & $6.50 \times 10^6$ & $7.57 \times 10^{-6}$ \\
  $p p \to t ~\bar{t}~W^\pm$ & $4.29 \times 10^1$ & $7.00 \times 10^6$ & $6.13 \times 10^{-6}$ \\
  $p p \to t ~\bar{t}~Z$ & $5.35 \times 10^1$ & $7.0 \times 10^6$ & $7.64 \times 10^{-6}$ \\
  $p p \to t ~\bar{t} + {\rm jets}$ & $5.36 \times 10^4$ & $7.38 \times 10^7$ & $7.26 \times 10^{-4}$ \\
  $p p \to t ~j + {\rm h.c.}$ & $3.65 \times 10^4$ & $2.85 \times 10^7$ & $1.28 \times 10^{-3}$ \\
  $p p \to t ~W^- + {\rm h.c.}$ & $6.02 \times 10^3$ & $1.55 \times 10^7$ & $3.88 \times 10^{-4}$ \\
  \toprule
  Signal~($p p \to t~H$) & $9.15 \times 10^{-1}$ & $6.0 \times 10^6$ & $1.53 \times 10^{-7}$ \\
  Signal~($p p \to t~\bar{t}$) & $5.39 \times 10^{0}$ & $8.0 \times 10^6$ & $6.74 \times 10^{-7}$ \\
  \toprule
  \end{tabular}
  \end{center}
    \caption{Total cross sections at $\sqrt{s} = 13.6$~TeV for the considered signal and background processes, computed at LO in QCD, together with the number of generated events $N_\mathrm{events}$ and the per-event weight $\omega=\sigma_{\rm LO}/N_{\rm events}$. The background rates are inclusive, while the signal entries include the $\mathrm{BR}(t\to b\ell\nu)$ and $\mathrm{BR}(H\to b\bar{b})$ branching ratios, together with the $\mathrm{BR}(t\to qH)$ one for pair-production, and are evaluated for the benchmark scenario $g_{qt}^{v}=g_{qt}^{a}=0.02$. In this case, associated production includes both the $u$- and $c$-initiated channels and their charge conjugates, while the pair-production entry uses $t\to uH$ (with a numerically indistinguishable rate for the $t\to cH$ case).\label{tab:xs}}
\end{table}

\begin{table}[!t]
  \centering
  \setlength{\tabcolsep}{7pt}
  \renewcommand{\arraystretch}{1.25}
  \begin{tabular}{l *{2}{cc}}
  \toprule
   & \multicolumn{2}{c}{Backgrounds} & \multicolumn{2}{c}{Signal ($tH$)} \\
  \cmidrule(lr){2-3}\cmidrule(lr){4-5}
  Cut & Events & $\varepsilon$ [\%] & Events & $\varepsilon$ [\%] \\
  \midrule
  Initial & $1.34\times 10^{7}$ & --- & $127.2$ & --- \\
  $N_{\ell}=1$ & $(6.926167\pm 0.000651)\times 10^{6}$ & 51.76 & $68.671\pm 0.026$ & 53.99 \\
  $\tau$ veto & $(6.476375\pm 0.000651)\times 10^{6}$ & 48.40 & $62.942\pm 0.026$ & 49.49 \\
  $E_{\mathrm{T}}^{\mathrm{miss}}>30~\mathrm{GeV}$ & $(4.958386\pm 0.000628)\times 10^{6}$ & 37.06 & $49.556\pm 0.025$ & 38.96 \\
  $N_{\mathrm{jets}}\ge 3$, $N_{b}\ge 3$ & $(3.42138\pm 0.00184)\times 10^{5}$ & 2.56 & $17.996\pm 0.018$ & 14.15 \\
  \bottomrule
  \end{tabular}
  \caption{Expected event yields and cumulative efficiencies after the common preselection for the summed backgrounds and the associated-production signal at $\mathcal{L}=139~\mathrm{fb}^{-1}$, using the rates and benchmark scenario of table~\ref{tab:xs}. Efficiencies are defined relative to the initial yields, and the quoted uncertainties are the MC statistical errors. \label{tab:cutflow_th}}
  \vspace{.625cm}\setlength{\tabcolsep}{12pt}
  \begin{tabular}{l *{2}{cc}}
  \toprule
   & \multicolumn{2}{c}{Signal ($t\bar{t}\to Hu$)} & \multicolumn{2}{c}{Signal ($t\bar{t}\to Hc$)} \\
  \cmidrule(lr){2-3}\cmidrule(lr){4-5}
  Cut & Events & $\varepsilon$ [\%] & Events & $\varepsilon$ [\%] \\
  \midrule
  Initial & $749.2$ & --- & $749.2$ & --- \\
  $N_{\ell}=1$ & $399.102\pm 0.132$ & 53.27 & $399.326\pm 0.132$ & 53.30 \\
  $\tau$ veto & $361.321\pm 0.132$ & 48.23 & $360.916\pm 0.132$ & 48.17 \\
  $E_{\mathrm{T}}^{\mathrm{miss}}>30~\mathrm{GeV}$ & $281.016\pm 0.128$ & 37.51 & $280.661\pm 0.128$ & 37.46 \\
  $N_{\mathrm{jets}}\ge 3$, $N_{b}\ge 3$ & $131.860\pm 0.101$ & 17.60 & $140.853\pm 0.103$ & 18.80 \\
  \bottomrule
  \end{tabular}
  \caption{Same as table~\ref{tab:cutflow_th} for the $t\bar{t}$ FCNC signals
   $t\to uH$ and $t\to cH$.\label{tab:cutflow_tt}}
\end{table}

\subsection{Event selection and reconstruction}
\label{sec:event_selection}

We consider separate analyses targeting associated $tH$ production and $t\bar t$ production with an FCNC top-quark decay, using a common simulation strategy and preselection. This is followed by reconstruction procedures adapted to the two topologies, which makes the corresponding signal regions mutually exclusive after the full event selection. For both analyses, we generate signal and background events with \textsf{MadGraph5\_aMC@NLO} version 3.4.2, following the setup described in~\cref{sec:lagrangian}, and use \textsf{MadWidth}~\cite{Alwall:2014bza} to compute decay widths and \textsf{MadSpin}~\cite{Artoisenet:2012st} to handle the decays of the top quark, Higgs boson and massive gauge bosons while preserving spin correlations. The decayed events are then passed to \textsf{Pythia}  8.3~\cite{Bierlich:2022pfr} for parton showering, hadronisation and hadron decays. The detector response is simulated with \textsf{Delphes}~\cite{deFavereau:2013fsa} version 3.5.1, with the particle efficiencies and resolutions described in refs.~\cite{Frank:2023epx, Esmail:2024gdc, Fuks:2025qgh} and including jet clustering with \textsf{FastJet} version 3.4.2~\cite{Cacciari:2011ma}, using the anti-$k_t$ algorithm~\cite{Cacciari:2008gp} with a radius parameter $R=0.4$.

Both analyses use a common preselection inspired by the ATLAS search for $t\to Hq$ decays of ref.~\cite{ATLAS:2018jqi}. Reconstructed electrons and muons are required to have a transverse momentum $\pT>25$~GeV and a pseudorapidity $|\eta|<2.4$, with the electrons in the transition region between the barrel and the end-cap calorimeters ($1.44<|\eta|<1.57$) being rejected. Both electrons and muons are further required to be isolated. To this aim, we define a track isolation parameter
\begin{equation}
  I_\ell = \sum_{\substack{i\,\in\,\mathrm{tracks}\\ \dR(i,\ell)<0.3}} \pT^{\,i}\,,
\end{equation}
as the scalar sum of the transverse momentum of the charged tracks inside a cone of radius $0.3$ around the lepton, excluding the lepton track itself, and accept a lepton only if $I_\ell/\pT^{\ell}<0.1$. Jets are selected if they satisfy $\pT>25$~GeV and $|\eta|<2.5$, and hadronically decaying $\tau$ leptons if $\pT>15$~GeV. Overlaps between leptons and jets are then removed sequentially: electrons and muons are first discarded if they lie within $\dR(e,j)<0.2$ and $\dR(\mu,j)<0.4$ of any selected jet, respectively, after which jets within $\dR(j,\ell)<0.4$ of a remaining lepton are discarded. We require events to contain exactly one signal lepton, no $\tau$-tagged jets, $E_T^{\rm miss}>30\,\mathrm{GeV}$, at least three jets and at least three $b$-tagged jets. The yields and cumulative efficiencies after this common preselection are shown in \cref{tab:cutflow_th} and \cref{tab:cutflow_tt} for an integrated luminosity of $\mathcal{L}=139~\mathrm{fb}^{-1}$. This preselection retains about 14\% of the single-production signal events and 17.6\% (18.8\%) of the pair-production signal events with $t\to uH$ ($t\to cH$), while only 2.6\% of the background events survives. The slightly higher efficiency for the pair-production signal with $t\to cH$ arises because the charm jet from the FCNC decay is occasionally identified as a $b$-jet.

For events passing the preselection, we first reconstruct the leptonically decaying top quark using the same procedure in both analyses. The leptonic $W$ boson is reconstructed from the charged lepton and the missing transverse momentum by imposing the $W$-mass constraint 
\begin{equation}
  \Big[m(p_\ell + p_\nu)\Big]^2 = M_W^2,
\label{eq:w_mass}
\end{equation}
where $p_\ell$ and $p_\nu$ are the four-momenta of the charged lepton and the neutrino, respectively, and $M_W=80.4\,\mathrm{GeV}$ is the $W$ boson mass. Identifying the neutrino transverse momentum with $\mathbf{p}_T^{\mathrm{miss}} = (p_{\rm miss, x}, p_{\rm miss, y})$, \cref{eq:w_mass} gives a quadratic equation for the neutrino longitudinal momentum $p_z^\nu$. When its discriminant is non-negative, we retain the solution with the smaller absolute value; otherwise, we use the real part of the complex solutions. The neutrino four-momentum is then $\bigl(E_\nu, p_{\rm miss,x},p_{\rm miss,y},p_z^\nu\bigr)$ with $E_\nu=\sqrt{p_{\rm miss,x}^2+p_{\rm miss,y}^2+(p_z^\nu)^2}$, and the leptonic top candidate is formed as
\begin{equation}
  t_{\ell} \;=\; \ell + \nu + b_{\ell}.
\end{equation}

The assignment of the $b$-tagged jets to the leptonic top and Higgs candidates is performed by minimising a $\chi^2$ discriminator over all triplets of selected $b$-jets and their permutations. Denoting by $b_1$ the $b$-jet associated with the leptonic top decay and by $b_2$ and $b_3$ those originating from the Higgs decay, with $p_H=p_{b_2}+p_{b_3}$ and $\dR_{bb}=\dR(b_2,b_3)$, the common $\chi^2$ reads 
\begin{equation}
  \chi^2_{\mathrm{common}} = \left(\frac{m_H^{\mathrm{reco}}-m_H}{20\,\mathrm{GeV}}\right)^{\!2} + \left(\frac{m_{t_{\ell}}^{\mathrm{reco}}-m_t}{40\,\mathrm{GeV}}\right)^{\!2} + \left(\frac{\dR_{bb}-\dR_{bb}^{\mathrm{exp}}}{2}\right)^{\!2},
  \label{eq:chi2common}
\end{equation}
where $m_H=125\,\mathrm{GeV}$ and $m_t=173\,\mathrm{GeV}$. In the single-production topology, only three $b$-jets enter the fit and one takes $\dR_{bb}^{\mathrm{exp}}=2$, appropriate for a relatively soft Higgs. 
\begin{figure}[!t]
  \centering
  \includegraphics[width=0.495\linewidth]{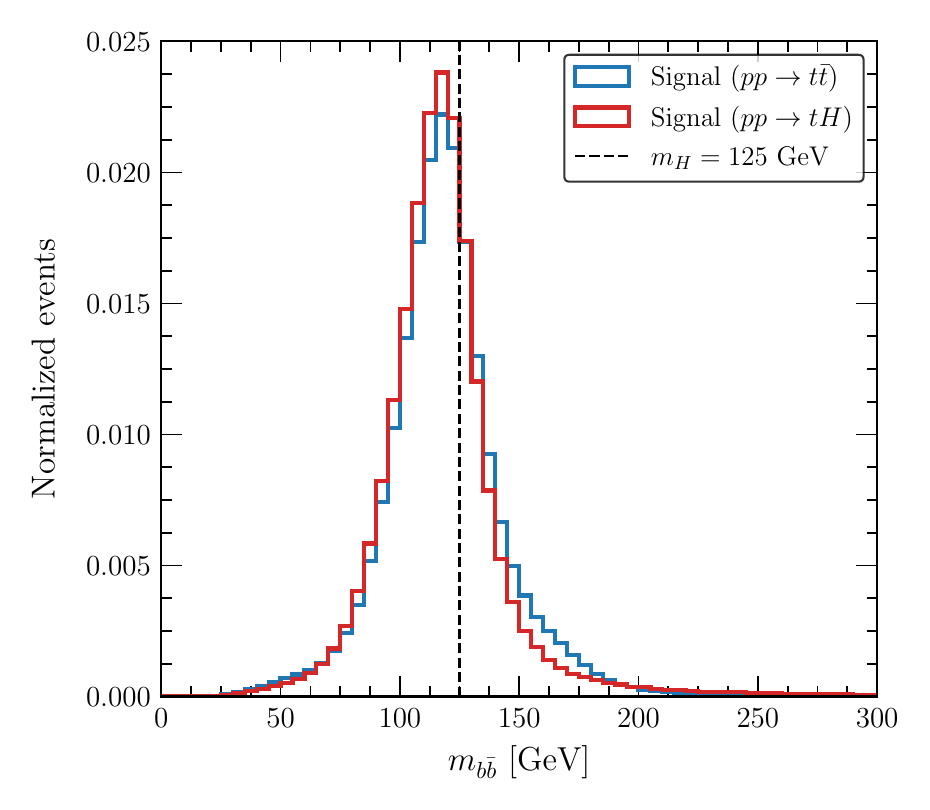}
  \hfill
  \includegraphics[width=0.495\linewidth]{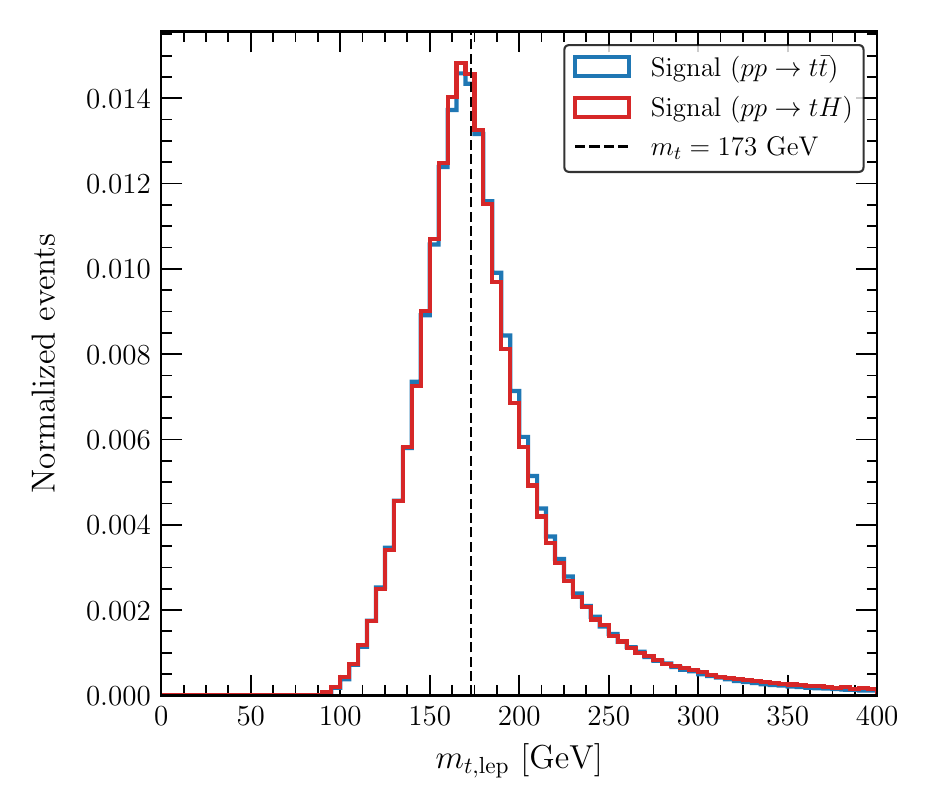}
  \vfill
  \includegraphics[width=0.495\linewidth]{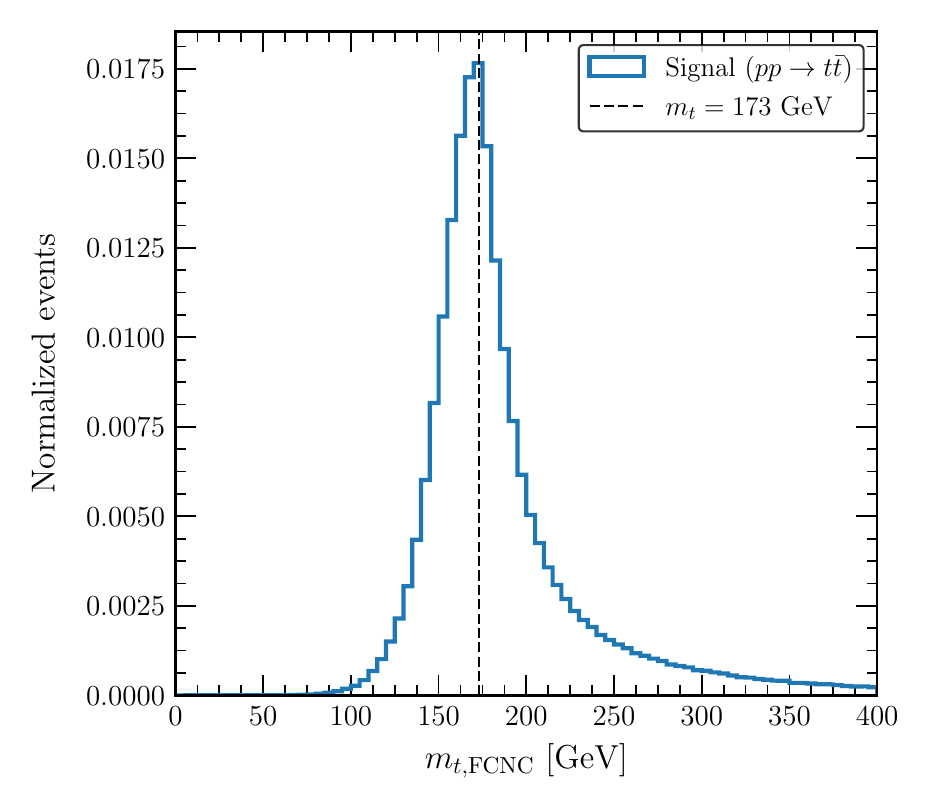}
  \caption{Normalised distributions of the masses of the reconstructed Higgs candidate $m_{b\bar{b}}$ (upper left), leptonic top $m_{t,\mathrm{lep}}$ (upper right) and FCNC top $m_{t,\mathrm{FCNC}}$ (lower) after the $\chi^2$ assignment. The blue and red curves correspond to pair and associated production, respectively, with the lower panel applying only to the former. The vertical dashed lines mark $m_H=125\,\mathrm{GeV}$ and $m_t=173\,\mathrm{GeV}$.}
  \label{fig:masses_reco}
\end{figure}
In the pair production topology, the FCNC decay $t\to qH$ produces an extra light quark recoiling against the Higgs boson. Events are therefore required to contain at least four jets (including the three $b$-jets), and the fit is extended by an additional jet candidate $q$ drawn from the remaining jets, after excluding $\{b_1,b_2,b_3\}$. The FCNC top is reconstructed as 
\begin{equation}
  t_{\mathrm{FCNC}} \;=\; q + H \;=\; q + b_2 + b_3,
\end{equation}
and the full discriminator becomes
\begin{equation}
  \chi^2 = \chi^2_{\mathrm{common}} + \left(\frac{m_{t_{\mathrm{FCNC}}}^{\mathrm{reco}}-m_t}{40\,\mathrm{GeV}}\right)^{\!2} + \delta_{b\text{-tag}},
\label{eq:chi2_pair}\end{equation}
where the penalty $\delta_{b\text{-tag}}=1$ ($0$) if the quark jet candidate is (is not) $b$-tagged, thereby favouring candidates that are not $b$-tagged. Because the Higgs boson originating from the $t\to qH$ decay is typically more boosted than the one produced in the associated-production channel, the expected $b\bar{b}$ opening angle constraint in \cref{eq:chi2common} is taken to be $\pT$-dependent. In the Higgs rest frame the two $b$-jets from the $H\to b\bar{b}$ decay are back-to-back so that under a large transverse boost they become collimated, with a characteristic angular separation of order $\Delta R_{bb} \approx 2m_H/\pT^{H}$. We therefore use this value as $\dR_{bb}^{\mathrm{exp}}$, restricting it to the window $[0.4,\,3.5]$ so that extreme values of $\pT^{H}$ do not drive the $\chi^2$ term pathologically. Events for which no valid assignment is found or, in the pair production mode, for which no eligible additional jet candidate remains, are discarded.

\Cref{fig:masses_reco} shows the normalised distributions of the reconstructed Higgs and top masses after the $\chi^2$ assignment for a benchmark scenario $g_{qt}^{v}=g_{qt}^{a}=0.02$. Both the pair production signal and the single production signal exhibit a clear peak in $m_{b\bar{b}}$ near the Higgs mass and in $m_{t,\mathrm{lep}}$ near the top mass, whereas in the pair production channel, the FCNC top mass $m_{t,\mathrm{FCNC}}=m(q+H)$ likewise peaks near $m_t$, as expected from the additional constraint in eq.~\eqref{eq:chi2_pair}. The two $m_{b\bar{b}}$ distributions are in good agreement, with small differences reflecting the distinct kinematics and reconstruction prescriptions of the two analyses.

\section{Graph neural network classifiers}
\label{sec:ml}
\begin{figure}
   \centering
    \includegraphics[width=.9\linewidth]{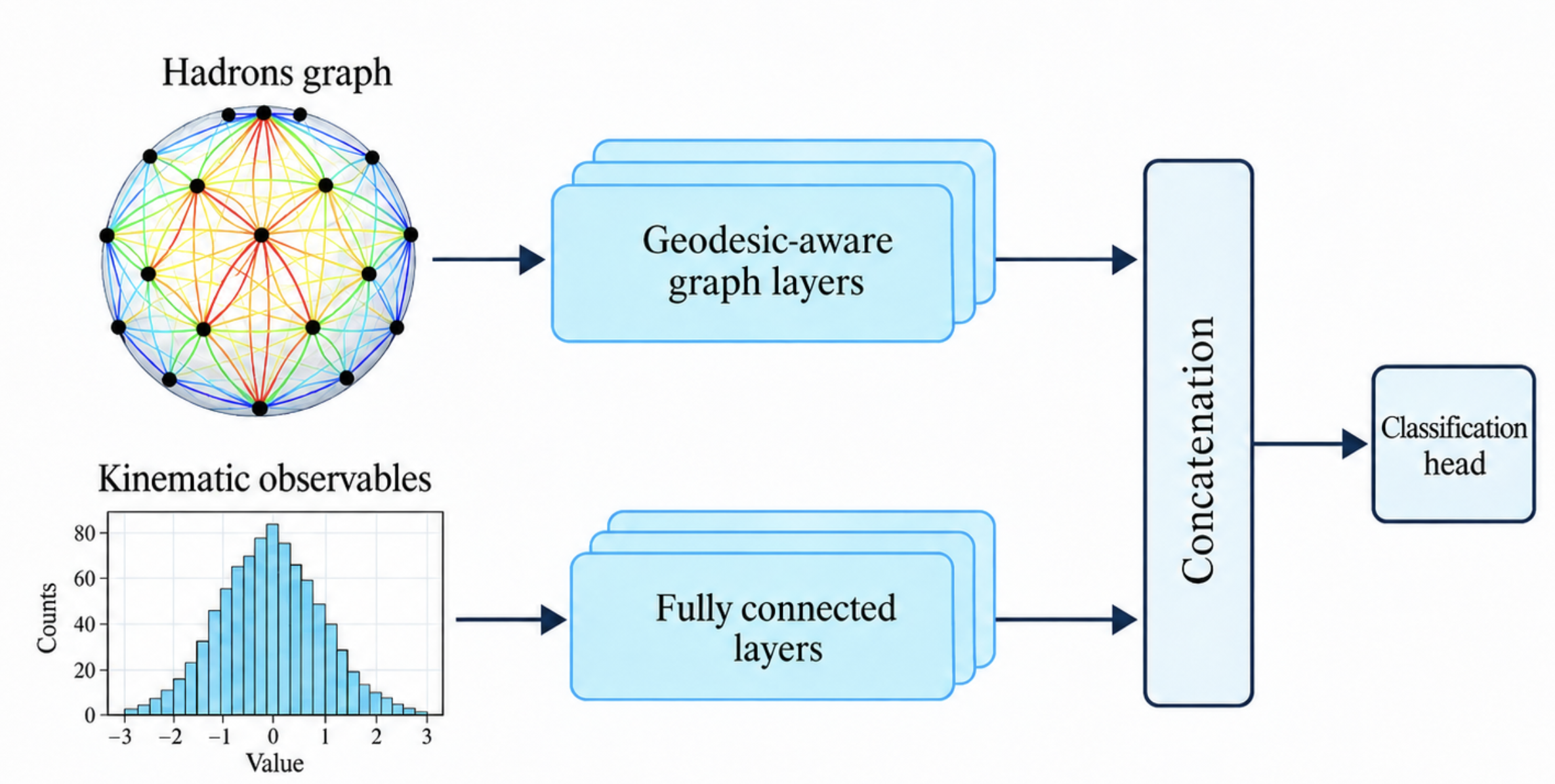}
    \vspace{-0.5cm}
    \caption{Architecture of the multimodal classifiers, combining a GNN branch for the hadronic constituents with an MLP branch for high-level kinematic observables. The spherical graph illustrates the GA-GNN, which uses geodesic distances, whereas the plain GNN uses a graph in the $(\eta,\phi)$ plane with $\dR$ distances.  The outputs of the GNN and MLP branches are concatenated and passed to a classification head that produces a signal score.}
    \label{fig:network}
\end{figure}

To exploit the differences between the hadronic radiation patterns of signal and background together with the reconstructed event kinematics, we compare two multimodal classifiers, referred to as the GA-GNN and the plain GNN, whose common design is illustrated in \cref{fig:network}. In both cases, a GNN branch processes the hadronic constituents of each event and an MLP branch processes high-level kinematic observables, with their outputs being combined by a classification head to distinguish signal from background. The two branches are trained jointly, allowing each encoder to adapt its representation to the complementary information provided by the other~\cite{Hammad:2023sbd, Hammad:2024hhm}. The two classifiers differ in the inputs and graph construction of their GNN branch: the GA-GNN uses the spherical representation introduced in~\cref{sec:colourflow} with geodesic distances, whereas the plain GNN operates in the $(\eta,\phi)$ plane with $\dR$ distances.

\subsection{Inputs and graph construction}
\label{sec:graphs}

For each event passing the selection described in section~\ref{sec:event_selection}, the kinematic branch receives a high level feature  vector $\mathbf{x}_{\rm kin}$ built from the reconstructed objects and event observables. It comprises the four-momenta of the charged lepton, the three assigned $b$-jets and the reconstructed Higgs boson and leptonic top quark, as well as those of up to five additional jets not assigned by the $\chi^2$ reconstruction\footnote{In the pair-production analysis, the jet candidate assigned to the FCNC top quark is included in this set.}, together with their masses. It also contains pairwise $\dR$ and $\dphi$ separations among the reconstructed objects, together with their azimuthal separations from the missing transverse momentum, as well as $E_T^{\rm miss}$, $m_T^{W}$, $H_T$, $m_{\mathrm{eff}}$, the lepton charge and the jet multiplicity. Here $m_T^W=\sqrt{2\pT^\ell E_T^{\rm miss} [1-\cos\Delta\phi(\ell,\mathbf{p}_T^{\rm miss})]}$ is the transverse mass of the leptonic $W$ candidate, with $\Delta\phi(\ell,\mathbf{p}_T^{\rm miss})$ denoting the azimuthal separation between the charged lepton and the missing transverse momentum. The scalar sum of the transverse momenta of all selected jets defines $H_T=\sum_j\pT^j$, while the effective mass $m_{\rm eff}=H_T+\pT^\ell+E_T^{\rm miss}$ also includes the charged lepton and the missing transverse momentum. 

The GNN branch receives the $n_\mathrm{real}$ charged and neutral hadronic constituents reconstructed by \textsf{Delphes} with $\pT>0.5$~GeV, ordered by decreasing transverse momentum and truncated to the first $250$ constituents when necessary, with zero padding applied to events containing fewer constituents. Each event is represented as a graph $\mathcal{G}=(\mathcal{V},\mathcal{E})$ whose nodes $\mathcal{V}$ correspond to the retained hadronic constituents, excluding the zero-padded entries. 

For the GA-GNN, the node features are 
\begin{equation}
  \mathbf{x}_i = \left(\pT,\eta,\phi,q,E,\theta,\phi_s,h\right)_i\in\mathbb{R}^{8}\,,
  \label{eq:nodes}
\end{equation}
where $\eta$ and $\phi$ denote the original detector coordinates, while $(\theta,\phi_s)$ are obtained from the centred coordinates through the spherical mapping, with all azimuthal and polar angles expressed in radians. Moreover, $q\in\{-1,0,+1\}$ is the hadron electric charge, $E$ its energy, $p_T$ its transverse momentum, and $h=+1$ ($-1$) identifies the northern (southern) hemisphere. The edges $\mathcal{E}$ are built with the $k$-nearest-neighbour ($k$NN) algorithm, using the geodesic (great-circle) distance on $S^2$,
\begin{equation}
  d_{\rm geo}(i,j) = \arccos\!\left( \cos\theta_i\cos\theta_j + \sin\theta_i\sin\theta_j\cos(\phi_{s,i}-\phi_{s,j}) \right).
  \label{eq:geodesic}
\end{equation}
This intrinsic distance is the length of the shorter great circle arc connecting the two nodes $i$ and $j$, rather than the straight-line chord through the embedding Euclidean space $\mathbb{R}^3$. It correctly accounts for the periodicity of $\phi_s$ and is invariant under a common rotation of all points on the sphere. The graph construction is summarised below.

\begin{algobox}{Graph construction}
\begin{enumerate}
  \setlength{\itemsep}{0.35em}
  \item Remove zero-padded nodes (all features identically zero).
  \item Compute the full $n_{\rm real}\times n_{\rm real}$ pairwise geodesic distance matrix $D_{ij} = d_{\rm geo}(i,j)$; set $D_{ii}=+\infty$ to exclude self-loops.
  \item For each node $i$, identify the $k_{\rm eff} = \min(k,\, n_{\rm real}-1)$ nearest neighbours by geodesic distance ($k=15$ by default).
  \item Add directed edges $(j\to i)$ for each neighbour $j$ of $i$, with edge attribute $e_{ij} = d_{\rm geo}(i,j)\in\mathbb{R}^1$.
\end{enumerate}
\vspace{0.15em}
\noindent For very sparse events with $n_{\rm real} < k+1$, $k_{\rm eff}$ is capped so that every node is connected to all remaining real nodes.
\end{algobox}

The plain GNN follows the same graph-construction procedure in the $(\eta,\phi)$ plane, using the $d_{\dR}$ distance to select the neighbouring nodes and as the edge attribute,
\begin{equation}
  d_{\dR}(i,j) = \sqrt{(\eta_i-\eta_j)^2 + (\dphi_{ij})^2}\,,
  \label{eq:dR}
\end{equation}
where $\dphi_{ij} = [(\phi_i-\phi_j+\pi)\bmod 2\pi]-\pi$ is the wrapped azimuthal difference. Its node features are the first five components of eq.~\eqref{eq:nodes}, $(\pT,\eta,\phi,q,E)_i\in\mathbb{R}^5$, without the sphere coordinates and the hemisphere flag. The comparison of the performance of the GA-GNN and plain GNN classifiers thus probes the combined effect of the spherical node features and the geodesic graph construction.

\subsection{Network architecture and training}
\label{sec:arch}

In both classifiers, the graphs defined above are processed by three edge conditioned convolution (NNConv) layers~\cite{Gilmer:2017mpnn, SimonovskyK17}, in which the messages exchanged between neighbouring nodes depend on the scalar edge attribute $e_{ij}$ given by the geodesic distance $d_{\rm geo}$ or $d_{\dR}$. For each directed edge $(j\to i)$, a small edge MLP produces a weight matrix
\begin{equation} 
  \Theta_{ij} = \mathrm{MLP}_{\rm edge}(e_{ij})
  \in\mathbb{R}^{C_{\rm out}\times C_{\rm in}}\,,
  \label{eq:nnconv_edge}
\end{equation}
where $C_{\rm in}$ and $C_{\rm out}$ denote the input and output feature dimensions of the layer. This matrix transforms the features of the neighbouring node $j$, and the resulting messages are averaged over the neighbours of node $i$. We denote by $\mathbf{h}_i^{(l)}$ the representation of node $i$ after layer $l$, with $\mathbf{h}_i^{(0)}=\mathbf{x}_i$. The three layers have hidden dimension $C=128$, ELU activations and a dropout rate of $0.2$. After the last layer $L=3$, the node representations are averaged over the nodes of the event to obtain a fixed-size representation of the hadronic radiation, 
\begin{equation}
  \mathbf{z}_{\GNN} = \frac{1}{|\mathcal{V}|}\sum_{i\in\mathcal{V}}
  \mathbf{h}_i^{(L)} \in \mathbb{R}^C\,.
  \label{eq:pool}
\end{equation}
The kinematic branch similarly maps $\mathbf{x}_{\rm kin}$ to a learned representation $\mathbf{z}_{\rm kin}\in\mathbb{R}^{D_{\rm kin}}$ with $D_{\rm kin}=64$ through fully connected layers. Combining this vector with the $C=128$ components of $\mathbf{z}_{\rm GNN}$ then gives the concatenated event representation,
\begin{equation}
  \mathbf{z}_{\rm fused} =
  \left[\mathbf{z}_{\GNN}\,\|\,\mathbf{z}_{\rm kin}\right]
  \in \mathbb{R}^{C+D_{\rm kin}}\,,
  \label{eq:fusion}
\end{equation}
which is passed through a classification head producing a single logit $\hat{y}$. The classifier assigns a signal score $\hat{p} = \sigma(\hat{y})$, where $\sigma$ is the sigmoid function.

Both architectures are trained separately for each of the three signal channels, namely associated $tH$ production and $t\bar t$ production with either a $t\to uH$ or a $t\to cH$ decay. For each channel, the dataset is divided into training, validation and test samples using a shuffled event-index permutation with a fixed random seed,
\begin{equation}
  \text{Training: } 60\%,
  \qquad
  \text{Validation: } 15\%,
  \qquad
  \text{Test: } 25\%.
  \label{eq:split}
\end{equation}
The same permutation is applied to the particle-level arrays, the kinematic vectors and the labels. The networks are trained by minimising the binary cross-entropy, using the settings  summarised in table~\ref{tab:hyper}. Both classifiers use the same optimiser, learning-rate schedule and early stopping criterion, and the checkpoint with the largest area under the receiver operating characteristic curve (ROC AUC) on the validation sample is retained. Their batch sizes, gradient accumulation and numerical precision differ as specified in the table.

\begin{table}[!p]
  \centering
  \small
  \setlength{\tabcolsep}{5pt}
  \renewcommand{\arraystretch}{1.15}
  \begin{tabular}{>{\raggedright\arraybackslash}p{3.9cm} >{\raggedright\arraybackslash}p{4.9cm} >{\raggedright\arraybackslash}p{4.9cm}}
  \toprule
  Hyperparameter & plain GNN & GA-GNN \\
  \midrule
  \multicolumn{3}{l}{\emph{Data and graph construction}} \\
  \midrule
  Total sample & \multicolumn{2}{>{\raggedright\arraybackslash}p{10.15cm}}{$5\times10^{5}$ signal $+$ $5\times10^{5}$ background ($10^{6}$ events, balanced)} \\
  Test sample & \multicolumn{2}{>{\raggedright\arraybackslash}p{10.15cm}}{$25\%$ of the total ($2.5\times10^{5}$ events)} \\
  Training / validation & \multicolumn{2}{>{\raggedright\arraybackslash}p{10.15cm}}{remaining $75\%$ split $80/20$ ($6\times10^{5}$ / $1.5\times10^{5}$ events)} \\
  Node features & $(\pT,\eta,\phi,q,E)$ & $(\pT,\eta,\phi,q,E,\theta,\phi_s,h)$ \\
  Edge metric & $d_{\Delta R}(i,j)$ in the $(\eta,\phi)$ plane, with $\Delta\phi\in[-\pi,\pi]$, eq.~\eqref{eq:dR} & geodesic distance $d_{\rm geo}(i,j)$ on the unit sphere, eq.~\eqref{eq:geodesic} \\
  Graph connectivity & \multicolumn{2}{>{\raggedright\arraybackslash}p{10.15cm}}{$k$-nearest neighbours in the edge metric, $k=15$, no self-loops} \\
  Edge-attribute dimension & \multicolumn{2}{>{\raggedright\arraybackslash}p{10.15cm}}{$1$ (the distance $d_{\Delta R}(i,j)$ or $d_{\rm geo}(i, j)$)} \\
  Zero-padded nodes & \multicolumn{2}{>{\raggedright\arraybackslash}p{10.15cm}}{removed before graph construction} \\
  Event-level inputs & \multicolumn{2}{>{\raggedright\arraybackslash}p{10.15cm}}{standardised with the training-set mean and standard deviation} \\
  \midrule
  \multicolumn{3}{l}{\emph{Architecture (NNConv)}} \\
  \midrule
  Message-passing layers & \multicolumn{2}{>{\raggedright\arraybackslash}p{10.15cm}}{$3\times$ NNConv (edge-conditioned), mean aggregation} \\
  Hidden dimension & \multicolumn{2}{>{\raggedright\arraybackslash}p{10.15cm}}{$128$} \\
  Edge network & \multicolumn{2}{>{\raggedright\arraybackslash}p{10.15cm}}{Linear($1\to16$) -- GELU -- Linear($16\to C_{\rm in}C_{\rm out}$)} \\
  Activation & \multicolumn{2}{>{\raggedright\arraybackslash}p{10.15cm}}{ELU} \\
  Graph readout & \multicolumn{2}{>{\raggedright\arraybackslash}p{10.15cm}}{global mean pooling} \\
  Kinematic MLP branch & \multicolumn{2}{>{\raggedright\arraybackslash}p{10.15cm}}{Linear($n_{\rm kin}\to256$) -- Linear($256\to128$) -- Linear($128\to64$), ELU, dropout} \\
  Kinematic embedding size & \multicolumn{2}{>{\raggedright\arraybackslash}p{10.15cm}}{$64$} \\
  Classification head & \multicolumn{2}{>{\raggedright\arraybackslash}p{10.15cm}}{Linear($192\to32$) -- ELU -- dropout -- Linear($32\to1$)} \\
  Dropout rate & \multicolumn{2}{>{\raggedright\arraybackslash}p{10.15cm}}{$0.2$} \\
  Output / decision & \multicolumn{2}{>{\raggedright\arraybackslash}p{10.15cm}}{logit $\to$ sigmoid, threshold $0.5$} \\
  \midrule
  \multicolumn{3}{l}{\emph{Training}} \\
  \midrule
  Loss function & \multicolumn{2}{>{\raggedright\arraybackslash}p{10.15cm}}{binary cross-entropy with logits} \\
  Optimiser & \multicolumn{2}{>{\raggedright\arraybackslash}p{10.15cm}}{AdamW ($\beta_1=0.9$, $\beta_2=0.999$, $\epsilon=10^{-8}$)} \\
  Initial learning rate & \multicolumn{2}{>{\raggedright\arraybackslash}p{10.15cm}}{$3\times10^{-4}$} \\
  Weight decay & \multicolumn{2}{>{\raggedright\arraybackslash}p{10.15cm}}{$10^{-4}$} \\
  Learning-rate schedule & \multicolumn{2}{>{\raggedright\arraybackslash}p{10.15cm}}{cosine annealing per epoch, $T_{\max}=50$, $\eta_{\min}=10^{-6}$} \\
  Batch size (per step) & $64$ & $128$ \\
  Gradient accumulation & $4$ steps & $1$ step \\
  Effective batch size & $256$ (per GPU) & $128$ \\
  Maximum epochs & \multicolumn{2}{>{\raggedright\arraybackslash}p{10.15cm}}{$50$} \\
  Early stopping & \multicolumn{2}{>{\raggedright\arraybackslash}p{10.15cm}}{validation AUC, patience of $5$ epochs} \\
  Model selection & \multicolumn{2}{>{\raggedright\arraybackslash}p{10.15cm}}{checkpoint with the best validation AUC} \\
  Numerical precision & FP16 mixed precision (loss in FP32) & FP32 \\
  Hardware & all allocated GPUs (DDP if $>1$) & $1$ GPU \\
  \bottomrule
  \end{tabular}
  \caption{Input specifications, network architecture and training settings of the plain GNN and the GA-GNN. The sample sizes refer to each signal channel, and entries spanning both columns apply to both classifiers. Here, FP32  and FP16 denote single- and half-precision floating-point arithmetic, respectively, and DDP denotes distributed data-parallel training. 
  \label{tab:hyper}}
\end{table}

Both the GA-GNN and the plain GNN models are implemented in \textsf{PyTorch}~\cite{paszke2019pytorch}, with \textsf{PyTorch Geometric}~\cite{fey2019fast} providing the graph operations, edge-conditioned convolutions and pooling, while \textsf{PyTorch Lightning}~\cite{falcon2019pytorch} manages training, checkpointing and early stopping. The \textsf{Scikit-Learn} library~\cite{pedregosa2011scikit} is used for data preprocessing and the computation of evaluation metrics.

\subsection{Classifier performance}
\label{sec:performance}

\begin{figure}[!t]
  \centering
  \includegraphics[width=\linewidth]{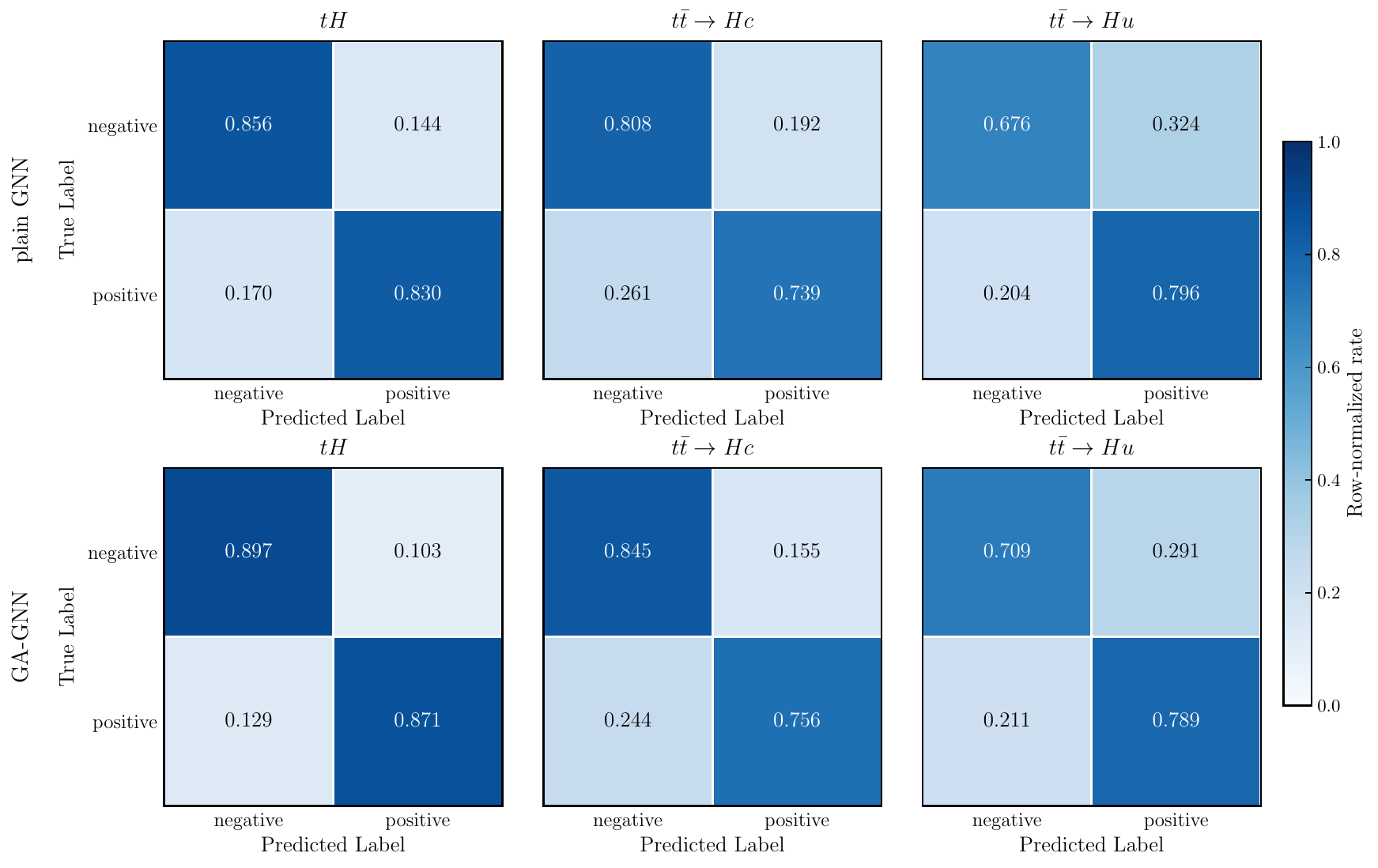}
  \caption{Confusion matrices for the plain GNN (upper panels) and the GA-GNN (lower panels), evaluated on the test samples at a score threshold of $0.5$. The left panels correspond to associated $tH$ production, while the central and right panels show $t\bar t$ production with $t\to cH$ and $t\to uH$, respectively. Each entry gives the fraction of events of a given true class assigned to the indicated predicted class, with ``negative'' denoting background and ``positive'' denoting signal.
  \label{fig:confusion}}
\end{figure}

The performance of the two classifiers is evaluated on the test samples for the three signal channels, with \cref{fig:confusion} showing the row-normalised confusion matrices at a score threshold of $0.5$. At this threshold, the GA-GNN reduces the fraction of background events classified as signal (false positive rate) relative to the plain GNN, while retaining a comparable or larger fraction of signal events (true positive rate). For associated $tH$ production, the false positive rate falls from $0.144$ to $0.103$ and the true positive rate rises from $0.830$ to $0.871$. For $t\bar t$ production with $t\to Hc$, the false positive rate falls from $0.192$ to $0.155$ while the true positive rate rises from $0.739$ to $0.756$, and for $t\bar t$ production with $t\to Hu$, the false positive rate falls from $0.324$ to $0.291$ at an almost unchanged true positive rate ($0.796$ and $0.789$).

\begin{figure}[!htb]
  \centering
  \includegraphics[width=0.46\linewidth]{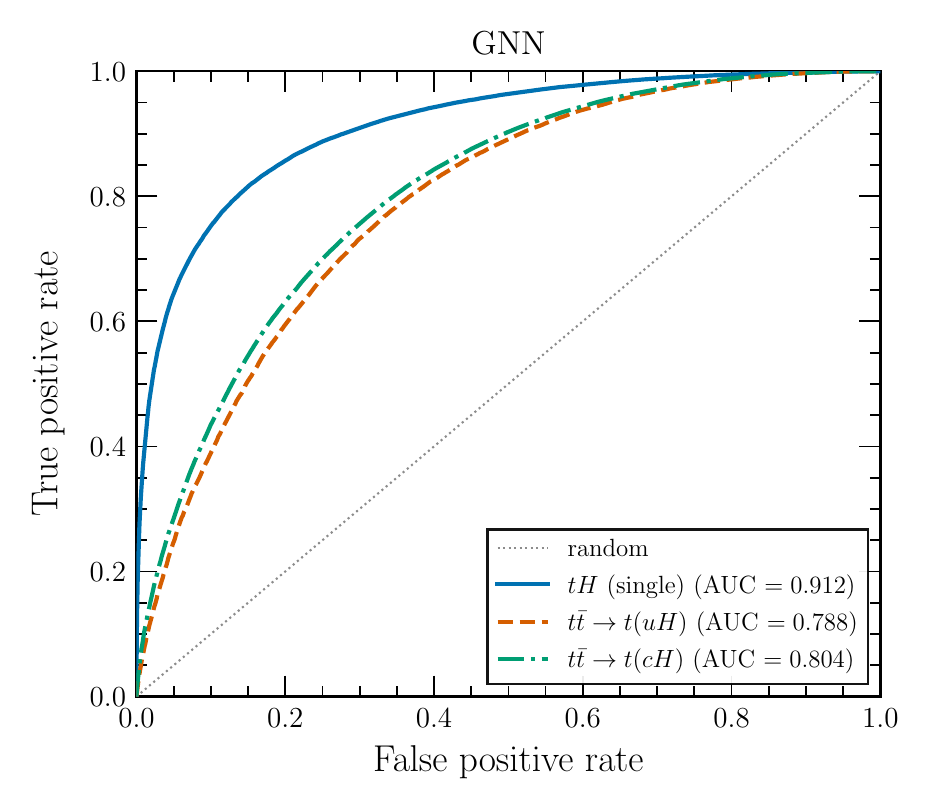}
  \hfill
  \includegraphics[width=0.46\linewidth]{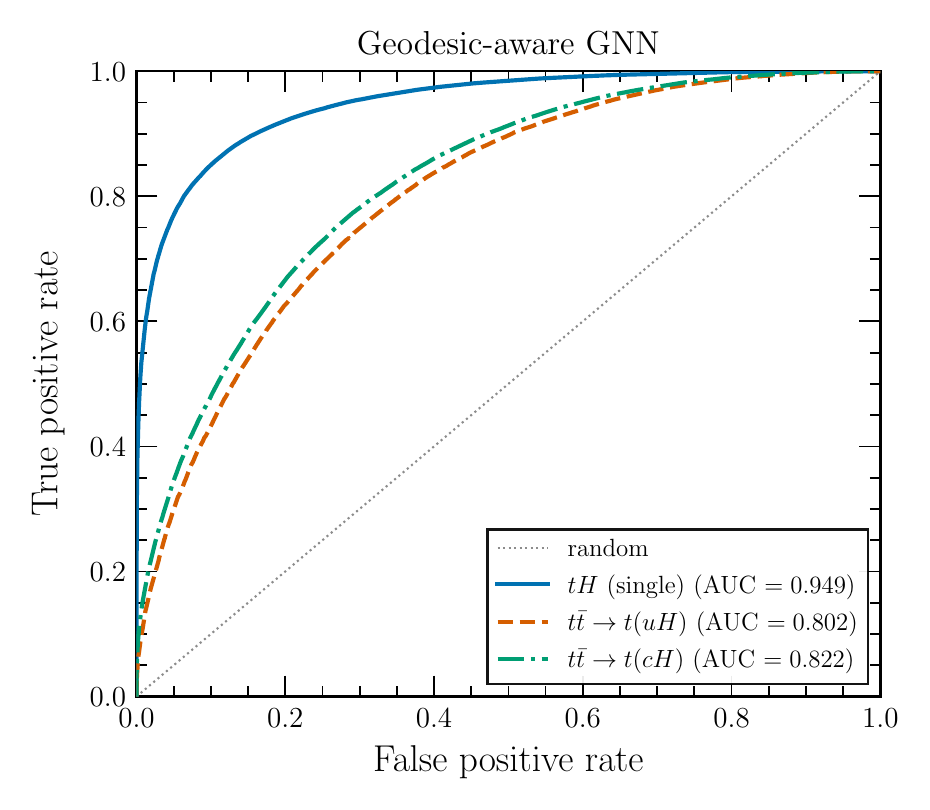}
  \caption{ROC curves evaluated on the test samples for the plain GNN (left) and the GA-GNN (right), for associated $tH$ production and $t\bar t$ production followed by a $t\to uH$ or $t\to cH$ decay. The corresponding AUC values are given in the legends, and the diagonal dotted line indicates random classification.
  \label{fig:roc_GNN}}
\end{figure}

The ROC curves of \cref{fig:roc_GNN} extend this comparison across classifier thresholds, showing how the retained signal efficiency varies with the fraction of background events accepted. The AUC increases from $0.912$ to $0.949$ for associated production, from $0.804$ to $0.822$ for $t\bar t$ production with $t\to Hc$ and from $0.788$ to $0.802$ for $t\bar t$ production with $t \to Hu$. The lower AUCs in the pair-production channels are consistent with their closer resemblance to the dominant $t\bar t+\mathrm{jets}$ background and with the additional quark jet originating from $t\to qH$, which can complicate the event reconstruction and contribute to radiation around the Higgs candidate. While the AUC provides a global measure of discrimination, the sensitivity to the considered rare signals depends particularly on the signal efficiency retained at very small background acceptance. We therefore assess the impact of the classifiers through the expected exclusion limits in section~\ref{sec:results}, taking into account the expected signal and background yields and their uncertainties.

\section{Statistical interpretation and expected limits}
\label{sec:results}

Having characterised the discrimination achieved by the two classifiers, we now translate their outputs into expected exclusion limits on the top FCNC couplings and the branching fractions $\mathrm{BR}(t\to qH)$, with $q=u,c$, as a function of the integrated luminosity~$\mathcal{L}$. We evaluate the sensitivity under the background-only hypothesis using three statistical frameworks that we introduce in \cref{sec:stat}. Framework A provides an asymptotic (Asimov) counting estimate with a single fractional background uncertainty, restricted to working points with sufficiently large expected background yields, while  framework B uses an exact Poisson treatment suitable for small event counts. Our main results are obtained with framework C, a profile-likelihood analysis incorporating a more detailed uncertainty model, which also allows us to assess the robustness of the sensitivity estimates obtained with the simpler treatments. All three frameworks describe each signal region as a single-bin counting experiment and report median expected upper limits at 95\% confidence level (CL), using the modified frequentist $\mathrm{CL}_s$ prescription~\cite{Junk:1999kv, Read:2002hq}. General reviews of the methods used below can be found in refs.~\cite{Cranmer:2015nia, ParticleDataGroup:2024cfk}, and recent discussions of likelihood-based inference and its implementation in collider reinterpretation studies can also be found in refs.~\cite{Alguero:2022gwm, Araz:2023bwx, Araz:2025bww}. The resulting expected limits are presented in section~\ref{sec:limits}, while \cref{sec:discussion} discusses their implications and the limitations of the analysis.

\subsection{Statistical methods}
\label{sec:stat}

\subsubsection{Event yields and classifier working points}
\label{sec:yields}

After the selection described in section~\ref{sec:event_selection}, events are retained if the classifier score exceeds a threshold~$c$, and the expected signal and background yields $s(c)$ and $b(c)$ can be written as
\begin{equation}
  s(c) = \mathcal{L}\,\sigma_S\,\varepsilon_S^{\rm pre}\,\varepsilon_{\rm tag}\,\mathrm{TPR}(c)\,,
  \qquad
  b(c) = \mathcal{L}\,\varepsilon_{\rm tag}\,\sum_k \mathrm{FPR}_k(c) \sigma_k\,\varepsilon_k^{\rm pre} \equiv \sum_k b_k(c)\,.
\end{equation}
Here, $\sigma_S$ denotes the signal cross section including the relevant decay branching fractions, $\sigma_k$ is the cross section of background process $k$, and the sum runs over all background contributions. The efficiencies $\varepsilon^{\rm pre}_{S,k}$ account for the event selection preceding the classifier cut, with the triple-$b$-tagging efficiency used in the detector simulation factored out and replaced by the explicit factor $\varepsilon_{\rm tag}=\varepsilon_b^3$ with $\varepsilon_b$ being the per-jet $b$-tagging efficiency. The subsequent classifier cut retains a fraction $\mathrm{TPR}(c)$ of the signal and $\mathrm{FPR}_k(c)$ of the background process $k$. As mentioned above, all background processes are combined with their cross-section weights into a single background class for the training and validation of the classifiers, so the combined sample is characterised by an overall false-positive rate, averaged over the individual processes with weights proportional to their expected yields before the classifier cut.

Both signal processes are proportional to the same combination of couplings. At leading order $\sigma(qg\to tH)\propto g_{qt}^2$, with $g_{qt}^2\equiv|g_{qt}^{v}|^2+|g_{qt}^{a}|^2$, while the pair-production rate is proportional to $\mathrm{BR}(t\to qH)$, which by eq.~\eqref{eq:width} is also proportional to $g_{qt}^2$ up to the negligible change of $\Gamma_t$. We therefore express all results in terms of the signal-strength modifier $\mu = \sigma_S/\sigma_S^{0}$, defined with respect to the rate $\sigma_S^0$ predicted for a reference coupling $g_{qt}^{0}=0.1$. An upper limit $\mu_{\rm up}$ translates into
\begin{equation}
  g_{qt} < g_{qt}^{0}\,\sqrt{\mu_{\rm up}}\,,
  \qquad
  \mathrm{BR}(t\to qH) < \mu_{\rm up}\;\mathrm{BR}(t\to qH)\big|_{g_{qt}^0}\,.
  \label{eq:coupling}
\end{equation}
For each flavour hypothesis, the mutually exclusive single- and pair-production signal regions are combined through a likelihood containing independent Poisson counting terms with a common signal-strength parameter $\mu$, which is translated into a limit on $\mathrm{BR}(t\to qH)$. In framework~A, the background nuisance parameters are treated as independent between regions, so  the individual Asimov test statistics add and the combined exclusion condition reads
\begin{equation}
  \sum_i Z_{{\rm excl},i}^2\geq \big[\Phi^{-1}(1-\alpha/2)\big]^2\,,
\end{equation}
where $\Phi$ denotes the cumulative distribution function of the standard normal distribution and $\Phi^{-1}$ its inverse, \textit{i.e.}~the corresponding quantile function. We set $\alpha=0.05$~\cite{Cowan:2010js, ATLAS:2011tau}, for which $\Phi^{-1}(1-\alpha/2)=1.96$, as appropriate for the median expected 95\% CL upper limit in the asymptotic $\mathrm{CL}_s$ prescription. In frameworks~B and~C, the combined limits are obtained from the joint Poisson distribution and the joint profile likelihood, respectively, with the treatment of uncertainties specified below.

The finite size of the simulated test sample introduces an uncertainty on the background acceptance. For an unweighted background test sample containing $N_{\rm test}$ events, a classifier threshold retains $N_{\rm sel}=\mathrm{FPR}\times N_{\rm test}$ events, giving an approximate relative uncertainty $\sigma_{\rm MC}/b\simeq1/\sqrt{N_{\rm sel}}$ at small background acceptance. This uncertainty is added in quadrature to the background uncertainty in frameworks~A and~B, and is modelled by a dedicated nuisance parameter in framework~C. When no background test event survives, we replace the empirical zero acceptance by $\mathrm{FPR}=3/N_{\rm test}$, corresponding to the approximate one-sided 95\% upper bound, and assign a relative MC uncertainty of 100\%. 

With these yields and uncertainties defined, the classifier threshold is selected independently in each signal region: for the GA-GNN in frameworks~B and~C, it maximises $s/b$, yielding a working point independent of  the luminosity. In framework~A, the threshold instead minimises the median expected upper limit subject to $b\geq b_{\min}=10$, and can therefore vary. For the plain GNN, the $s/b$ optimum occurs in a region with a negligible signal yield, so all frameworks minimise their respective median  expected limits, with no background floor imposed in frameworks~B and~C. The false-positive rates and the expected yields at $139~\mathrm{fb}^{-1}$ are given in table~\ref{tab:wp}. At $\varepsilon_b=70\%$ the plain-GNN thresholds of frameworks~A and~B agree in the $t\bar{t}$ regions and differ in single production; at $\varepsilon_b=85\%$ they agree in both.

\begin{table}[!t]
  \centering
  \footnotesize
  \setlength{\tabcolsep}{6pt}
  \renewcommand{\arraystretch}{1.12}
  \begin{tabular}{@{}l cccc ccc ccc@{}}
    \toprule
    & \multicolumn{4}{c}{Associated $tH$ production}
    & \multicolumn{3}{c}{$t\bar{t}$, $t\to uH$}
    & \multicolumn{3}{c}{$t\bar{t}$, $t\to cH$} \\
    \cmidrule(lr){2-5} \cmidrule(lr){6-8} \cmidrule(lr){9-11}
    Framework & FPR & $s_u$ & $s_c$ & $b$
    & FPR & $s$ & $b$
    & FPR & $s$ & $b$ \\
    \midrule
    \multicolumn{11}{@{}l}{\emph{$\varepsilon_b=70\%$, GA-GNN}} \\
    A & $9.0\times10^{-5}$ & $15$ & $1.8$ & $11$
      & $4.2\times10^{-4}$ & $14$ & $49$
      & $9.0\times10^{-5}$ & $22$ & $11$ \\
    B, C & $1.0\times10^{-5}$ & $11$ & $1.3$ & $1.2$
         & $2.0\times10^{-5}$ & $3.0$ & $2.3$
         & $1.0\times10^{-5}$ & $14$ & $1.2$ \\
    \addlinespace
    \multicolumn{11}{@{}l}{\emph{$\varepsilon_b=70\%$, plain GNN}} \\
    A & $1.7\times10^{-4}$ & $4.1$ & $0.50$ & $20$
      & $5.1\times10^{-3}$ & $24$ & $6.0\times10^{2}$
      & $3.0\times10^{-3}$ & $26$ & $3.5\times10^{2}$ \\
    B & $3.6\times10^{-4}$ & $6.0$ & $0.74$ & $42$
      & $5.1\times10^{-3}$ & $24$ & $6.0\times10^{2}$
      & $3.0\times10^{-3}$ & $26$ & $3.5\times10^{2}$ \\
    C & $6.0\times10^{-5}$ & $2.5$ & $0.30$ & $7.0$
      & $1.7\times10^{-3}$ & $10$ & $2.0\times10^{2}$
      & $2.6\times10^{-4}$ & $3.8$ & $31$ \\
    \midrule
    \multicolumn{11}{@{}l}{\emph{$\varepsilon_b=85\%$, GA-GNN}} \\
    A & $5.0\times10^{-5}$ & $24$ & $3.0$ & $11$
      & $4.2\times10^{-4}$ & $26$ & $88$
      & $5.0\times10^{-5}$ & $35$ & $11$ \\
    B, C & $1.0\times10^{-5}$ & $19$ & $2.4$ & $2.1$
         & $2.0\times10^{-5}$ & $5.3$ & $4.2$
         & $1.0\times10^{-5}$ & $26$ & $2.1$ \\
    \addlinespace
    \multicolumn{11}{@{}l}{\emph{$\varepsilon_b=85\%$, plain GNN}} \\
    A, B & $6.0\times10^{-5}$ & $4.4$ & $0.54$ & $13$
         & $5.1\times10^{-3}$ & $44$ & $1.1\times10^{3}$
         & $2.3\times10^{-3}$ & $38$ & $4.8\times10^{2}$ \\
    C & $6.0\times10^{-5}$ & $4.4$ & $0.54$ & $13$
      & $1.7\times10^{-3}$ & $18$ & $3.6\times10^{2}$
      & $2.6\times10^{-4}$ & $6.8$ & $55$ \\
    \bottomrule
  \end{tabular}
  \caption{Classifier working points at $139~\mathrm{fb}^{-1}$. The quantities $s$ and $b$ are the expected signal and background yields at $g_{qt}^{0} \equiv \sqrt{(g_{qt}^v)^2 + (g_{qt}^a)^2}=0.1$, and $\varepsilon_b$ denotes the per-jet $b$-tagging efficiency. In associated production, the two flavours share the false-positive rate and background yield, while $s_u$ and $s_c$ denote the signal yields for the $u$- and $c$-initiated processes, respectively, including their charge conjugates. When two statistical frameworks are listed in the same row, they select the same working point in each signal region.
  \label{tab:wp}}
\end{table}

\subsubsection{Framework A: asymptotic counting estimate}
\label{sec:asimov}

In framework~A, the background is assigned a fractional normalisation uncertainty $\delta_b=0.10$ at $\mathcal{L}=139~\mathrm{fb}^{-1}$, which is combined in quadrature with the Monte Carlo statistical uncertainty, $\sigma_b^2 = (\delta_b\, b)^2 + \sigma_{\rm MC}^2$. This uncertainty is represented by an effective auxiliary Poisson measurement that constrains the background yield. Following refs.~\cite{Cowan:medsig, Cowan:2012talk}, we model the latter as a Poisson count $m\sim\mathrm{Pois}(\tau b)$ with $\tau=b/\sigma_b^2$, so that the likelihood reads as follows:
\begin{equation}
  \mathcal{L}(\mu,\beta) = \mathrm{Pois}(n\,|\,\mu s+\beta)\; \mathrm{Pois}(m\,|\,\tau\beta)\,,
  \label{eq:onoff}
\end{equation}
where $\beta$ is the background-yield nuisance parameter. Upper limits on $\mu$ are derived from the profile-likelihood-ratio test statistic $q_\mu$~\cite{Cowan:2010js}, and its median under the background-only hypothesis is approximated by evaluating it on the Asimov dataset, $n=b$ and $m=\tau b$, in which the observables are set to their expectation values. For $\mu=1$, this gives the following estimate of the median exclusion significance~\cite{Kumar:2015tna}
\begin{equation}
  Z_{\rm excl} = \left[\,2\left(s - b\ln\frac{b+s+x}{2b}
    - \frac{b^2}{\sigma_b^2}\ln\frac{b-s+x}{2b}\right)
    - (b+s-x)\left(1+\frac{b}{\sigma_b^2}\right)\right]^{1/2},
  \label{eq:zexcl}
\end{equation}
with
\begin{equation}
  x = \sqrt{(s+b)^2 - \frac{4\,s\,b\,\sigma_b^2}{b+\sigma_b^2}}\,.
\end{equation}
Equation~\eqref{eq:zexcl} gives $\sqrt{q_{1,\mathrm A}}$ for the likelihood of eq.~\eqref{eq:onoff}, where the subscript $\mathrm A$ denotes evaluation on the background-only Asimov dataset.

For a single signal region, the median expected 95\% CL exclusion criterion introduced in section~\ref{sec:yields} becomes
\begin{equation}
  Z_{\rm excl}(s,b,\sigma_b) \geq 1.96\,.
  \label{eq:zcrit}
\end{equation}
Solving this condition for the signal yield gives $s_{\rm up}$ and hence $\mu_{\rm up}=s_{\rm up}/s$. Since framework~A relies on the asymptotic distribution of $q_\mu$, we restrict its application to working points satisfying $b(c)\geq b_{\min}=10$, as specified in section~\ref{sec:yields}. In the limiting case of a known background, we have verified that the value of $s_{\rm up}$ obtained from eq.~\eqref{eq:zcrit} lies below the exact median $\mathrm{CL}_s$ limit of framework~B by at most 10\% for $b\geq10$ and by at most 5\% for $b\geq50$. The difference oscillates with $b$ because the Poisson median is discrete: it amounts to 6.7\% at $b=10$ and 3.1\% at $b=50$, and reaches 9.4\% at $b\simeq10.7$.

\subsubsection{Framework B: exact Poisson counting}
\label{sec:poisson}

To evaluate the sensitivity at working points with only a few expected background events, framework~B uses the discrete Poisson probabilities directly, avoiding the asymptotic approximation of framework~A. For a single signal region, the observed event count $n$ serves as the test statistic. Uncertainties on the signal and on the background yield are implemented through log-normal constraints, which guarantee positive yields,
\begin{equation}
  s(\theta_s) = s\,(1+\delta_s)^{\theta_s}\,,\qquad
  b(\theta_b) = b\,(1+\delta_{b,\mathrm{eff}})^{\theta_b}\,,\qquad
  \theta_{s},\,\theta_b \sim \mathcal{N}(0,1)\,.
\end{equation}
Here $\delta_s$ parametrises the signal-normalisation uncertainty and we set $\delta_s=0$ so that the signal normalisation is held fixed, while $\delta_{b,\mathrm{eff}}=\sigma_b/b$ includes the background normalisation and MC statistical uncertainties introduced above. The nuisance parameters are marginalised following the hybrid Bayesian-frequentist prescription of Cousins and Highland~\cite{Cousins:1991qz},
\begin{equation}
  P(n\,|\,\mu) = \int \mathrm{Pois}\big(n\,\big|\,\mu\, s(\theta_s) + b(\theta_b)\big)\,
  \phi(\theta_s)\,\phi(\theta_b)\,\mathrm{d}\theta_s\,\mathrm{d}\theta_b\,,
  \label{eq:ch}
\end{equation}
where $\phi$ denotes the standard normal density, and we evaluate the integral numerically with Gauss-Hermite quadrature. Upper limits are set with the $\mathrm{CL}_s$ method~\cite{Junk:1999kv,Read:2002hq},
\begin{equation}
  \mathrm{CL}_s(\mu;\,n_{\rm obs}) =
  \frac{P(n \le n_{\rm obs}\,|\,\mu)}{P(n \le n_{\rm obs}\,|\,0)}\,,
  \qquad
  \mathrm{CL}_s(\mu_{\rm up};\,n_{\rm obs}) = \alpha = 0.05\,.
  \label{eq:cls_count}
\end{equation}
For a known background and $n_{\rm obs}=0$, eq.~\eqref{eq:cls_count} gives the background-independent result $\mu_{\rm up}\,s=-\ln\alpha\simeq3$. For the combination of the two signal regions, we use the likelihood-ratio ordering statistic employed in the LEP Higgs searches~\cite{Junk:1999kv,Read:2002hq}, evaluated at the nominal signal and background yields,
\begin{equation}
  Q = n_1 \ln\!\left(1 + \mu\,\frac{s_1}{b_1}\right)
    + n_2 \ln\!\left(1 + \mu\,\frac{s_2}{b_2}\right),
\end{equation}
which reduces to ordering by $n$ for a single region. Its distribution under each hypothesis is summed exactly on the $(n_1,n_2)$ lattice, with the background nuisance parameters treated as independent between regions and marginalised as above, and the $\mathrm{CL}_s$ is constructed as in eq.~\eqref{eq:cls_count}, replacing the count-based tail probabilities by the corresponding probabilities $P(Q\le Q_{\rm obs}\,|\,\mu)$ and $P(Q\le Q_{\rm obs}\,|\,0)$.

For a single signal region, the expected limit and its uncertainty bands are defined through the distribution of $\mu_{\rm up}(n_{\rm obs})$ when $n_{\rm obs}$ is drawn from the background-only distribution $P(n\,|\,0)$. Since $\mu_{\rm up}(n_{\rm obs})$ increases monotonically with $n_{\rm obs}$, its quantiles are obtained by evaluating $\mu_{\rm up}$ at
the corresponding quantiles of $P(n\,|\,0)$ itself. We therefore evaluate the expected limits exactly, without pseudo-experiments. The median expected limit is $\mu_{\rm up}(n_{0.5})$, where $n_p$ is the smallest $n$ for which $P(n'\leq n\,|\,0)\geq p$. The $\pm1\sigma$ and $\pm2\sigma$ bands are obtained from the quantiles corresponding to $p\simeq0.1587$ and $0.8413$, and $p\simeq0.02275$ and $0.97725$, respectively. For the combined analysis, the corresponding quantiles are obtained from the distribution of $\mu_{\rm up}(n_1,n_2)$, weighting each pair of counts by its joint background-only probability. Owing to the discreteness of the event counts, the expected-limit curves can exhibit jumps as $\mathcal{L}$ varies, and different quantile curves may coincide. Framework~B also serves as a benchmark for framework~A, since both are built on the same counting experiment and the same exclusion criterion. Moreover, frameworks~B and~C share the log-normal form of the background constraint. As a validation check for a single signal region, retaining only a common background uncertainty $\delta_b=20\%$ and fixing all other nuisance parameters to their nominal values, the marginalised limits of B and the profiled limits of C agree to within 2\% for $b\geq50$.

\subsubsection{Framework C: profile-likelihood analysis}
\label{sec:profile}

Our main results are obtained with a profile-likelihood analysis incorporating separate nuisance parameters for the uncertainty sources described below. For a single signal region defined by a given classifier working point, the likelihood reads
\begin{equation}
  \mathcal{L}(\mu,\boldsymbol{\theta},\gamma) =
  \mathrm{Pois}\Big(n\,\Big|\,\mu\, s(\boldsymbol{\theta})
    + \gamma\sum_k b_k(\boldsymbol{\theta})\Big)\;
  \prod_j \mathcal{N}(0\,|\,\theta_j,1)\;
  \mathcal{N}\big(1\,\big|\,\gamma,\,\sigma_{\rm MC}/b\big)\,,
  \label{eq:likelihood}
\end{equation}
with
\begin{equation}
  s(\boldsymbol{\theta}) = s\prod_j (1+\delta_{S,j})^{\theta_j}\,,
  \qquad
  b_k(\boldsymbol{\theta}) = b_k\prod_j (1+\delta_{k,j})^{\theta_j}\,.
\end{equation}
The Poisson term describes the statistical fluctuations of the data, while each nuisance parameter $\theta_j$ is  associated with one source of systematic uncertainty, with relative impact $\delta_{S,j}$ on the signal and $\delta_{k,j}$ on the background process~$k$. The factors $(1+\delta)^{\theta_j}$ amount to a multiplicative log-normal response: the yield is unchanged at $\theta_j=0$, shifted by the relative amount $\delta$ at $\theta_j=+1$, and kept positive for all $\theta_j$, which consists in the normalisation systematic of the \textsc{HistFactory} statistical model~\cite{Cranmer:2012sba}. In the combined analysis, the likelihood contains one Poisson counting term for each signal region, with a common signal-strength parameter $\mu$. Experimental nuisance parameters are shared, and hence fully correlated, between the processes and signal regions they affect. Moreover, the background-normalisation uncertainty is represented by a single nuisance parameter acting on the total background yield within a signal region, and the background-normalisation and MC-statistical nuisance parameters are taken as independent between the two signal regions. The uncertainty sources and their adopted sizes are summarised in table~\ref{tab:syst}.

\begin{table}
  \centering
  \small\setlength\tabcolsep{5pt}\renewcommand{\arraystretch}{1.15}
  \begin{tabular}{l l c l}
    \toprule
    Source & Affects & Size & Ref.\\
    \midrule
    Background normalisation (all processes combined) & background & 50\% & \cite{ATLAS:2023mcc}\\
    \midrule
    Integrated luminosity & all & 1.7\% & \cite{ATLAS:2023mcc,ATLAS:2022hro}\\
    $b$-tagging efficiency ($3$ jets) & all & 5\% & \cite{ATLAS:2026bjet,ATLAS:2026bcjet}\\
    $c$/light mistag rates & background & 15\% & \cite{ATLAS:2026bcjet,ATLAS:2023lwk}\\
    Jet energy scale & all & 3\% & \cite{ATLAS:2020cli}\\
    Lepton identification, isolation and trigger & all & 1\% & \cite{ATLAS:2023mcc}\\
    \midrule
    Signal normalisation, $ug\to tH$ (band only) & signal & $^{+12.4}_{-10.8}\%$ & \cite{Alwall:2014hca,NNPDF:2021njg}\\
    Signal normalisation, $cg\to tH$ (band only) & signal & $^{+29.8}_{-29.5}\%$ & \cite{Alwall:2014hca,NNPDF:2021njg}\\
    Signal normalisation, $t\bar t$ with $t\to \,qH$ (band only) & signal & $^{+27.2}_{-20.1}\%$ & \cite{Alwall:2014hca,NNPDF:2021njg}\\
    MC statistics ($\gamma$) & background & $\sigma_{\rm MC}/b$ & \cite{Barlow:1993dm,Conway:2011in}\\
    \bottomrule
  \end{tabular}
  \caption{Uncertainty sources and adopted sizes for framework~C. The background-normalisation uncertainty applies to the total background yield, while the experimental uncertainties are correlated between the processes and signal regions they affect. The MC statistical uncertainty is implemented through a Gaussian constraint on the background scaling parameter $\gamma$, and signal-theory normalisation uncertainties are displayed as bands in the exclusion contours and are not included in the fit.\label{tab:syst}}
\end{table}

We use the inclusive false-positive rate of the combined background sample to determine the total background yield, to which we assign a single normalisation uncertainty of 50\%. After the requirement of three $b$-tagged jets, the background is dominated by $t\bar t$+jets production with additional heavy-flavour jets, which accounts for about 90\% of the selected background. Its normalisation is poorly known, and ATLAS assigns a 50\% normalisation uncertainty to the $t\bar t+{\geq}1b$ and $t\bar t+{\geq}1c$ components in the same final state~\cite{ATLAS:2023mcc}, substantially larger than the approximately $\pm4\%$ uncertainty of the inclusive NNLO+NNLL $t\bar t$ cross section~\cite{Czakon:2011xx}. Our $t\bar t$+jets sample is generated by merging matrix elements with up to two additional partons, providing a leading-order description of configurations with two extra hard jets. Although this extends the matrix-element description of additional radiation beyond that of the nominal inclusive $t\bar t$ sample used in ref.~\cite{ATLAS:2023mcc}, the precision of the inclusive NNLO+NNLL cross section cannot be assumed for the selected heavy-flavour component, and we therefore retain the uncertainty prescription of the ATLAS analysis. Furthermore, the scale and parton density uncertainties quoted for higher-order inclusive predictions of the subdominant $t\bar t X$ and single-top cross sections are typically at the level of a few percent to about 10\%, and thus smaller than the 50\% uncertainty adopted here~\cite{Karlberg:2024zps, Kulesza:2018tqz, Buonocore:2023ljm, LHCTopWG:SingleTop}. In our simplified treatment, these background uncertainties are therefore included in the common normalisation uncertainty assigned to the total background yield.

In addition to the background-normalisation uncertainty, we include assumed experimental uncertainties  informed by existing experimental studies, treating each source as correlated across the processes it affects. For the integrated luminosity, we adopt an uncertainty of 1.7\%, following ref.~\cite{ATLAS:2023mcc}, although the final ATLAS Run-2 calibration  reaches 0.83\%~\cite{ATLAS:2022hro}. For flavour tagging, calibrations of genuine $b$-jets at $\sqrt{s}=13.6$~TeV reach a precision of 1-2\% per jet~\cite{ATLAS:2026bjet,ATLAS:2026bcjet}, and we adopt a correlated uncertainty of 5\% on the event yields associated with the three tagged jets. The corresponding $c$-jet and light-jet mistag rates are known to better than 5\%~\cite{ATLAS:2026bcjet} and to 11-23\%~\cite{ATLAS:2023lwk}, respectively, and their impact, which mainly affects the background, is represented by an additional 15\% uncertainty on the background yields. Jet energy scale calibrations reach a precision of 1\% for central jets with $250~\mathrm{GeV} <\pT<2000$~GeV and 5\% at $\pT=20$~GeV~\cite{ATLAS:2020cli}; we represent their impact by a 3\% uncertainty on the event yields and include a further 1\% uncertainty associated with lepton identification, isolation and triggering, following ref.~\cite{ATLAS:2023mcc}.

We also evaluate the theoretical uncertainties on the signal normalisation, which we display as bands on the prediction without including them in the fit. The renormalisation-scale ($\mu_R$)  and factorisation-scale ($\mu_F$) uncertainties are estimated from the envelope of the seven-point variation of $\mu_R$ and $\mu_F$ by factors of two, while the parton-distribution-function (PDF) uncertainties are obtained from the replicas of the \textsf{NNPDF40} LO set~\cite{NNPDF:2021njg}. These relative uncertainties are computed with \textsf{MadGraph5\_aMC@NLO} and are independent of the overall FCNC coupling normalisation. For associated $tH$ production, they amount to $^{+12.1}_{-10.4}\%$ (scale) and $\pm2.9\%$ (PDF) for $ug\to tH$, and to $^{+11.5}_{-10.7}\%$ (scale) and $\pm27.5\%$ (PDF) for $cg\to tH$, with the larger PDF uncertainty in the latter case reflecting the poorly constrained charm content of the proton. For pair production, the rate is normalised to the NNLO+NNLL $t\bar t$ cross section, but we conservatively retain the relative uncertainties of the LO calculation, $^{+27.1}_{-20.0}\%$ (scale) and $\pm2.0\%$ (PDF). Combining the scale and PDF uncertainties in quadrature yields normalisation bands of $^{+12.4}_{-10.8}\%$ ($ug\to tH$), $^{+29.8}_{-29.5}\%$ ($cg\to tH$) and $^{+27.2}_{-20.1}\%$ ($t\bar t$ production with one $t\,\to qH$ decay).

Theory uncertainties affecting the signal and background acceptances are not included. Besides scale and PDF variations, they arise mainly from the modelling of the parton shower and hadronisation, to which the particle-level colour-flow information exploited by the classifier is directly sensitive~\cite{Sjostrand:2014zea}. This includes, in particular, colour reconnection, which can rearrange the colour connections between partons before hadronisation~\cite{Christiansen:2015yqa, Argyropoulos:2014zoa}. Although the Higgs decay products are not reconnected to the rest of the event in the default \textsf{Pythia} treatment~\cite{Bierlich:2022pfr}, colour reconnection elsewhere in the event can still affect the radiation patterns of both signal and background. Studies of particle-level QCD predictions have shown that uncertainties from parton showering and hadronisation can be sizeable~\cite{Amoroso:2018qga, Jueid:2022qjg, Jueid:2023vrb}, but a dedicated assessment of their impact on the present analysis is left for future work.

Finally, the finite size of the simulated background sample introduces a statistical uncertainty, which we include in the fit through the Barlow-Beeston \emph{lite} approach~\cite{Barlow:1993dm, Conway:2011in}, with a single positive parameter $\gamma$ that scales the total background and is constrained by a Gaussian term of relative width $\sigma_{\rm MC}/b$.

With the likelihood of \cref{eq:likelihood}, upper limits on $\mu$ are derived using the profile-likelihood-ratio test statistic appropriate for a non-negative signal strength~\cite{Cowan:2010js},
\begin{equation}
  \tilde{q}_\mu =
  \begin{cases}
    -2\ln\dfrac{\mathcal{L}\big(\mu,\hat{\hat{\boldsymbol{\theta}}}(\mu)\big)}
               {\mathcal{L}\big(0,\hat{\hat{\boldsymbol{\theta}}}(0)\big)}
      & \hat{\mu} < 0\,,\\[4ex]
    -2\ln\dfrac{\mathcal{L}\big(\mu,\hat{\hat{\boldsymbol{\theta}}}(\mu)\big)}
               {\mathcal{L}\big(\hat{\mu},\hat{\boldsymbol{\theta}}\big)}
      & 0 \le \hat{\mu} \le \mu\,,\\[4ex]
    0 & \hat{\mu} > \mu\,,
  \end{cases}
\end{equation}
where $\boldsymbol{\theta}$ denotes the full set of nuisance parameters, including $\gamma$, while $\hat{\mu}$ and $\hat{\boldsymbol{\theta}}$ maximise the likelihood and $\hat{\hat{\boldsymbol{\theta}}}(\mu)$ maximises it for fixed~$\mu$. Limits are set with the $\mathrm{CL}_s$ prescription~\cite{Junk:1999kv,Read:2002hq}, as in the LHC Higgs search procedure~\cite{ATLAS:2011tau}, and the asymptotic median expected limits are obtained from a background-only Asimov dataset, with the count in each signal region set to its nominal background expectation and the auxiliary measurements set to their nominal expectation values. In the asymptotic approximation, the median expected limit and the $\pm 1\sigma$ and $\pm2\sigma$ bands are given by~\cite{Cowan:2010js, Read:2002hq}
\begin{equation}
  \mu_{\rm up}^{(N)} = \sigma\left[\,N + \Phi^{-1}\big(1-\alpha\,\Phi(N)\big)\right],
  \quad\text{for}\quad N = 0,\pm1,\pm2
  \quad\text{and}\quad \sigma^2 = \frac{\mu^2}{\tilde q_{\mu,\rm A}}\,.
  \label{eq:bands}
\end{equation}
Here, $\alpha = 0.05$, $\tilde q_{\mu,\rm A}$ denotes the test statistic evaluated on the background-only Asimov dataset, and $\sigma$ is the corresponding estimate of the standard deviation of the signal-strength estimator $\hat{\mu}$. For $N=0$, this reproduces the median exclusion criterion  used in framework~A, $\sqrt{\tilde q_{\mu,\rm A}}=\Phi^{-1}(1-\alpha/2)=1.96$. In practice, the statistical model is implemented in the \textsf{HistFactory} format~\cite{Cranmer:2012sba} and evaluated with \textsf{pyhf}~\cite{Heinrich:2021gyp}, and the median expected limit and its bands are obtained by solving $\mathrm{CL}_s^{(N)}(\mu)=\alpha$, where $\mathrm{CL}_s^{(N)}$ denotes the corresponding quantile of the background-only distribution of $\mathrm{CL}_s$. Since the asymptotic approximation can become less reliable at small event counts, we validate its results with pseudo-experiments for working points with $b<b_{\min}$. At the GA-GNN working points used for the main results, where the background amounts to approximately one to two background events per signal region, the limits obtained from pseudo-experiments exceed the asymptotic ones by 8\% for  $\mathrm{BR}(t\to uH)$ and by 1\% for $\mathrm{BR}(t\to cH)$.

For comparison with framework~A, we consider a single signal region and retain only one background-normalisation nuisance parameter in framework~C, fixing all other nuisance parameters at their nominal values. When the background is known exactly, the two frameworks give identical median expected limits in the asymptotic $\mathrm{CL}_s$ treatment. Taking $\delta_b=20\%$ in both frameworks, the limits agree to within 2\% at $b=10$ and to within 9\% at $b=100$, with the residual difference arising from the auxiliary Poisson constraint in framework~A and the log-normal background response in framework~C.

\subsection{Expected limits and luminosity projections}
\label{sec:limits}
\begin{figure}[!t]
  \centering
  \includegraphics[width=\linewidth]{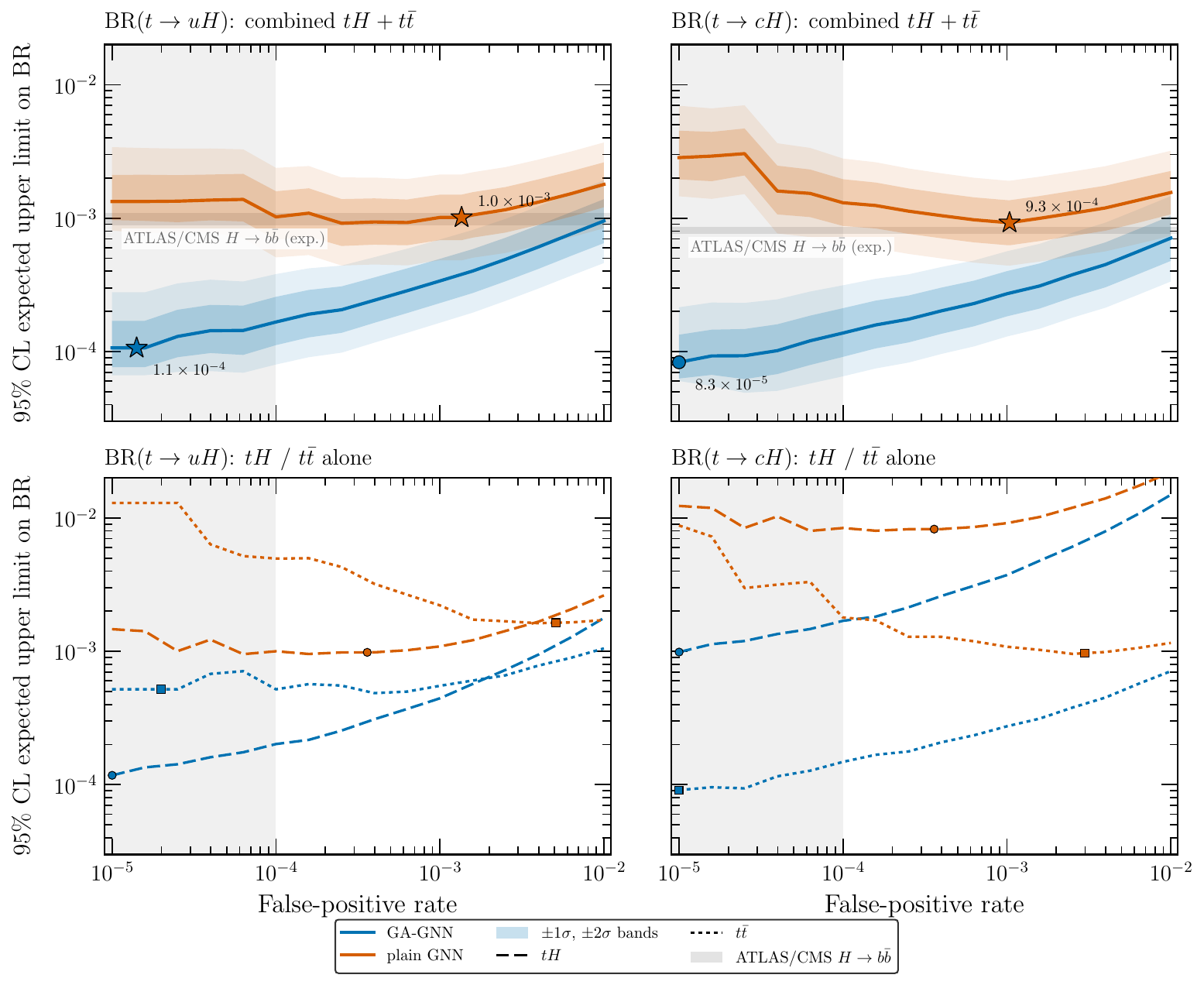}
  \caption{Median expected 95\%~CL upper limits on $\mathrm{BR}(t\to uH)$ (left) and $\mathrm{BR}(t\to cH)$ (right) as a function of the background FPR, obtained with framework~B at $139~\mathrm{fb}^{-1}$ and $\varepsilon_b=70\%$ for the GA-GNN (blue) and the plain GNN (orange). The upper panels combine the associated-production and pair-production signal regions at a common FPR, with the darker and lighter bands showing the $\pm1\sigma$/$\pm2\sigma$ expected ranges, respectively, under the background-only hypothesis, while the lower panels show the individual-region limits, with dashed curves for associated production and dotted curves for pair production. The circles and squares in the lower panels mark the framework~B working points listed in table~\ref{tab:wp} for associated and pair production, respectively, whereas the upper-panel markers show the combined limits reported in \cref{tab:limits139}, with their horizontal positions set to the geometric mean of the FPRs at the two regional working points. The grey horizontal bands indicate the range of the expected ATLAS and CMS limits in the $H\to b\bar{b}$ channel~\cite{ATLAS:2023mcc,CMS:2021ncc}, while the light vertical shading marks the region where fewer than about 10 background MC events survive.
  \label{fig:br_vs_wp}}
\end{figure}

We first illustrate the dependence of the expected branching fraction limits on the classifier working point, taking framework~B as an example. Fig.~\ref{fig:br_vs_wp} shows the median expected 95\%~CL upper limits on $\mathrm{BR}(t\to uH)$ and $\mathrm{BR}(t\to cH)$ as a function of the false positive rate (FPR) of the score cut, for an integrated luminosity of $139~\mathrm{fb}^{-1}$ and a fixed $b$-tagging efficiency $\varepsilon_b=70\%$. Two classifiers are compared throughout: the colour-flow GA-GNN and the plain GNN trained without the geometric embedding. The upper panels combine the associated $tH$ production and $t\bar{t}$ signal regions at a common FPR, with the shaded bands showing the $\pm1\sigma$ and $\pm2\sigma$ expected variations under the background-only hypothesis, while the lower panels show the corresponding limits from each region separately, and the grey horizontal strip marks the range of the expected ATLAS and CMS limits in the $H\to b\bar{b}$ channel at a comparable luminosity~\cite{ATLAS:2023mcc, CMS:2021ncc}. The shaded vertical band at $\mathrm{FPR}\lesssim10^{-4}$ flags working points with fewer than about ten background Monte Carlo events in the test sample, where the empirical FPR (and therefore the predicted yields) becomes MC-statistics limited.

The GA-GNN improves on the plain GNN over the full range of working points shown in \cref{fig:br_vs_wp}. At the working points specified in section~\ref{sec:yields} and listed in \cref{tab:wp}, framework~B yields combined median limits of $\mathrm{BR}(t\to uH)\simeq1.1\times10^{-4}$ and $\mathrm{BR}(t\to cH)\simeq8.3\times10^{-5}$ for the GA-GNN, roughly an order of magnitude stronger than the corresponding plain-GNN limits of $1.0\times10^{-3}$ and $9.3\times10^{-4}$. Associated production alone is competitive with the combination at the tightest cuts for the up-quark GA-GNN analysis, while pair production dominates the charm-quark sensitivity. \Cref{tab:limits139} collects the median expected limits of the three frameworks at $139~\mathrm{fb}^{-1}$ and $\varepsilon_b=70\%$. For the GA-GNN, frameworks~B and~C agree within 10\%, while framework~A is weaker because the requirement $b\ge 10$ excludes the highest-purity working points. For the plain GNN, the up-quark limits are comparable across the three frameworks, whereas the charm-quark limit is weaker in framework~C, which uses a larger background-normalisation uncertainty and selects a tighter $t\bar{t}$ threshold.

\begin{table}[!t]
  \centering
  \small
  \setlength{\tabcolsep}{7pt}
  \renewcommand{\arraystretch}{1.2}
  \begin{tabular}{@{}l ccc ccc@{}}
    \toprule
    & \multicolumn{3}{c}{$\mathrm{BR}(t\to uH)$}
    & \multicolumn{3}{c}{$\mathrm{BR}(t\to cH)$} \\
    \cmidrule(lr){2-4} \cmidrule(lr){5-7}
    Classifier & A & B & C & A & B & C \\
    \midrule
    GA-GNN    & $1.9\times10^{-4}$ & $1.1\times10^{-4}$ & $1.1\times10^{-4}$
              & $1.4\times10^{-4}$ & $8.3\times10^{-5}$ & $8.9\times10^{-5}$ \\
    \addlinespace
    plain GNN & $8.6\times10^{-4}$ & $1.0\times10^{-3}$ & $1.1\times10^{-3}$
              & $9.8\times10^{-4}$ & $9.3\times10^{-4}$ & $1.9\times10^{-3}$ \\
    \bottomrule
  \end{tabular}
  \caption{Median expected 95\%~CL upper limits on $\mathrm{BR}(t\to uH)$ and $\mathrm{BR}(t\to cH)$ at $139~\mathrm{fb}^{-1}$ and $\varepsilon_b=70\%$. For each flavour hypothesis, the mutually exclusive associated-production and pair-production signal regions are combined using a common signal-strength parameter, and the results are shown for both classifiers in frameworks~A, B and~C using their corresponding working points from table~\ref{tab:wp}.
  \label{tab:limits139}}
\end{table}

The greater sensitivity of the $t\bar{t}$ region to $t\to cH$ than to $t\to uH$ reflects differences in the signal efficiencies: for equal FCNC couplings, the two branching fractions are equal in the light-quark-mass approximation used here, while the background yield is the same at a fixed false-positive rate. At $\mathrm{FPR}=10^{-4}$, the true-positive rate for the $t\to cH$ pair-production signal is $3.1\times10^{-3}$ with the plain GNN and $3.7\times10^{-2}$ with the GA-GNN, compared with $1.3\times10^{-3}$ and $1.2\times10^{-2}$, respectively, for the $t\to uH$ pair-production case. Although the $H\to b\bar{b}$ system is common to both decays, the properties of the recoiling jet differ. First, the higher probability for a charm jet to be identified as a $b$-jet can affect both the event selection and the jet assignment. In our simulated samples, an additional $b$-tagged jet is indeed present in about 19\% of the selected pair-production events with $t\to cH$, compared with 14\% for $t\to uH$ and 10\% for the $t\bar{t}$+jets background. Second, differences in the mass and fragmentation of the recoiling quark jet may also contribute to the classifier discrimination. ATLAS and CMS likewise obtain somewhat stronger expected limits for $t\to cH$ in refs.~\cite{ATLAS:2023mcc, CMS:2021ncc}. However, while this efficiency difference is already present for the plain GNN and is larger for the GA-GNN, the background acceptance at this working point is estimated from only a small number of simulated events, and the resulting sensitivities should be confirmed with larger samples. The relative importance of the two flavours is reversed in associated production: for equal FCNC couplings, the leading order rate for $ug\to tH$ is about eight times that for $cg\to tH$, reflecting the larger up-quark density in the proton. For the plain GNN combination, this contribution partly compensates for the lower pair-production efficiency in the up-quark channel, with the relative ordering of the combined limits depending on the statistical framework. For the GA-GNN, the higher pair-production efficiency in the charm-quark channel instead leads to a stronger combined charm limit.

\begin{figure}[!t]
  \centering
  \includegraphics[width=\linewidth]{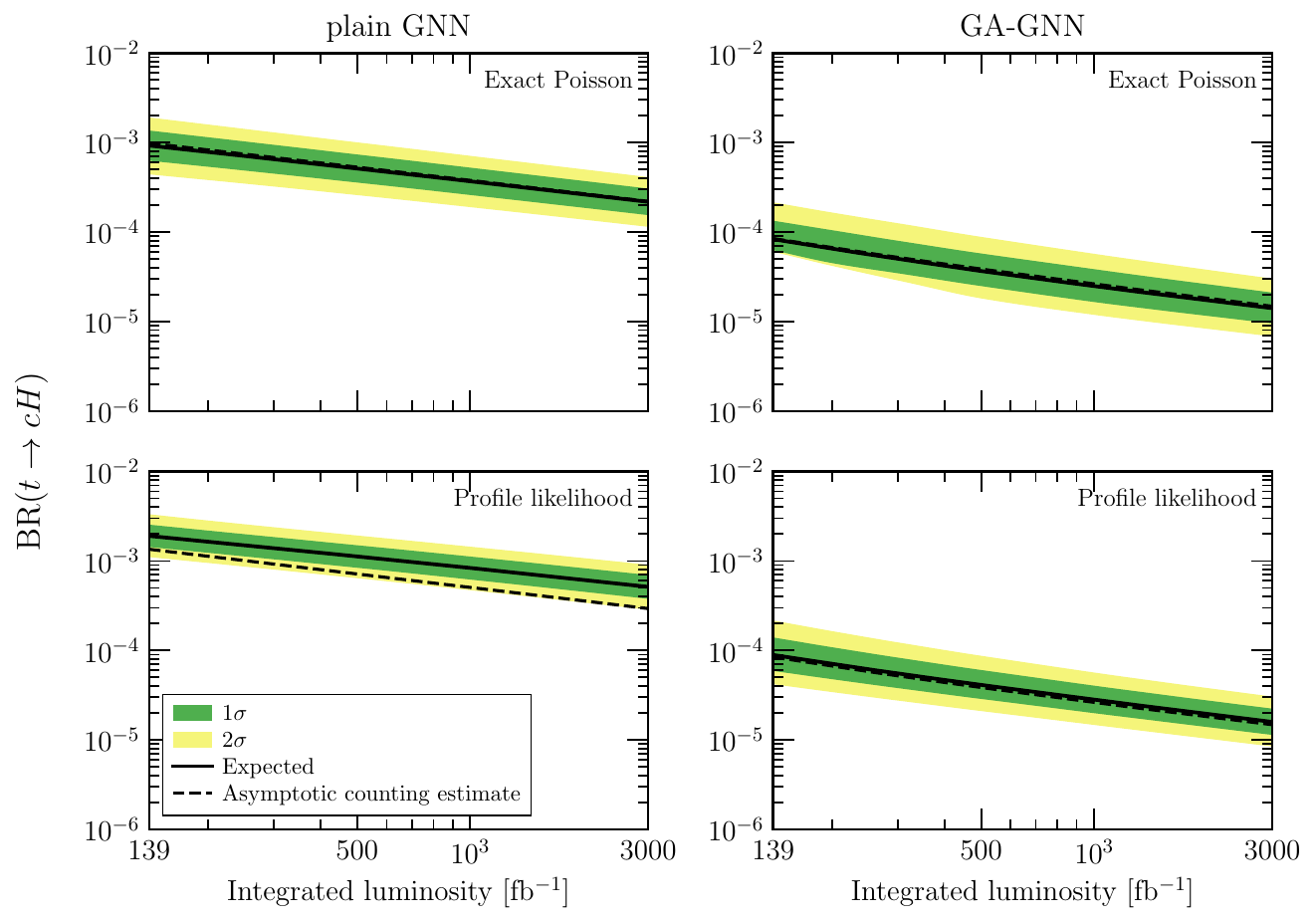}
  \caption{Median expected 95\%~CL upper limits on $\mathrm{BR}(t\to cH)$ as a function of the integrated luminosity for $\varepsilon_b=70\%$, combining the associated-production and pair-production signal regions. The upper panels show the results of framework~B and the lower panels those of framework~C, with the plain GNN on the left and the GA-GNN on the right. The solid black curves indicate the median expected limits, while the green and yellow bands show the $\pm1\sigma$ and $\pm2\sigma$ expected ranges under the background-only hypothesis. The dashed curves show the asymptotic counting estimate of eq.~\eqref{eq:zexcl}, evaluated at the working points used in each panel with the background-uncertainty prescription of framework~A without imposing its minimum-background requirement.
  \label{fig:br_lumi_c}}
\end{figure}

\begin{figure}[!t]
  \centering
  \includegraphics[width=\linewidth]{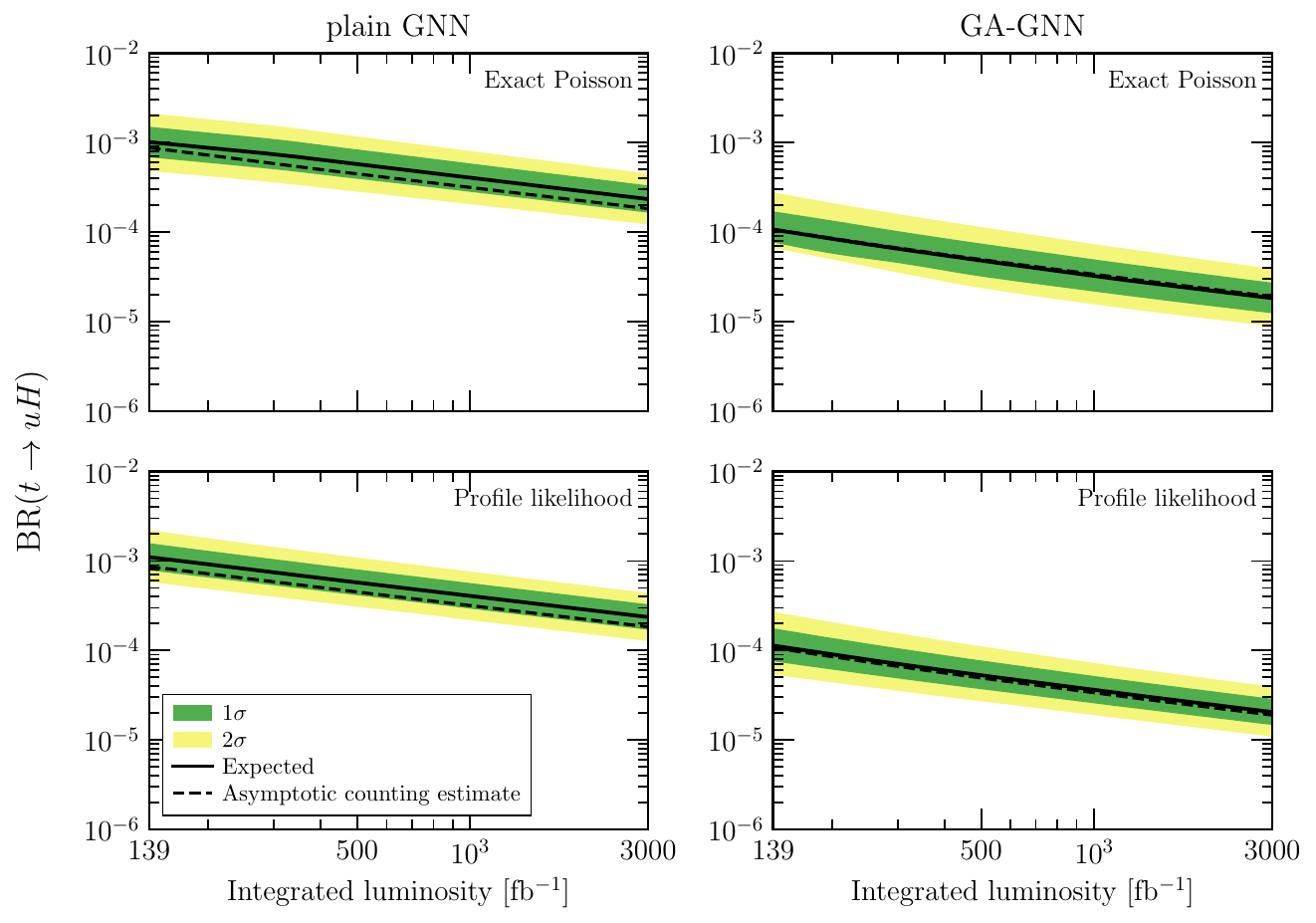}
  \caption{Same as \cref{fig:br_lumi_c}, but for $\mathrm{BR}(t\to uH)$.
  \label{fig:br_lumi_u}}
\end{figure}

We extend these sensitivity estimates to integrated luminosities between $139~\mathrm{fb}^{-1}$ and $3000~\mathrm{fb}^{-1}$ in \cref{fig:br_lumi_c,fig:br_lumi_u}. Signal and background yields are rescaled with the luminosity, while the classifier working points are selected according to the prescriptions described in \cref{sec:yields}.  For these projections, the relative uncertainties entering the statistical models are assumed to decrease as $\sqrt{139~\mathrm{fb}^{-1}/\mathcal{L}}$, which applies to the background-normalisation uncertainty $\delta_b$ in frameworks~A and~B, to the relative uncertainties associated with the nuisance parameters of framework~C including those affecting the signal, and to the MC statistical uncertainty in all three frameworks. The relative signal-theory normalisation uncertainties are instead kept fixed. This scaling assumes that background constraints and experimental calibrations improve with the available data, as discussed in \cref{sec:discussion}. Under these assumptions, our main profile-likelihood analysis (framework~C) with the GA-GNN reaches median expected limits of  $\mathrm{BR}(t\to uH)<2.1\times10^{-5}$ and $\mathrm{BR}(t\to cH)<1.6\times10^{-5}$ at $3000~\mathrm{fb}^{-1}$ and $\varepsilon_b=70\%$. For comparison, framework~B yields $\mathrm{BR}(t\to uH)<1.8\times10^{-5}$ and $\mathrm{BR}(t\to cH)<1.4\times10^{-5}$ for the same classifier and tagging efficiency, improving to $1.5\times10^{-5}$ and $1.2\times10^{-5}$, respectively, at $\varepsilon_b=85\%$, and for the plain GNN, framework~A at $3000~\mathrm{fb}^{-1}$ reaches $\mathrm{BR}(t\to uH)<1.8\times10^{-4}$ and $\mathrm{BR}(t\to cH)<2.2\times10^{-4}$ at $\varepsilon_b=70\%$, and $1.6\times10^{-4}$ and $2.1\times10^{-4}$ at $\varepsilon_b=85\%$. Finally, \cref{tab:br_lumi} collects the median expected limits of frameworks~B and~C at $300$ and $3000~\mathrm{fb}^{-1}$ for both tagging efficiencies, together with the $\pm2\sigma$ expected variations.

\begin{table}[!t]
  \centering
  \footnotesize
  \setlength{\tabcolsep}{8pt}
  \renewcommand{\arraystretch}{1.15}
  \begin{tabular}{@{}ll cccc@{}}
    \toprule
    & & \multicolumn{2}{c}{$\mathrm{BR}(t\to uH)$}
      & \multicolumn{2}{c}{$\mathrm{BR}(t\to cH)$} \\
    \cmidrule(lr){3-4} \cmidrule(lr){5-6}
    Framework & $\varepsilon_b$
      & $300~\mathrm{fb}^{-1}$ & $3000~\mathrm{fb}^{-1}$
      & $300~\mathrm{fb}^{-1}$ & $3000~\mathrm{fb}^{-1}$ \\
    \midrule
    \multicolumn{6}{@{}l}{\emph{GA-GNN}} \\
    B & $70\%$
      & $6.5^{+9.4}_{-3.0}\times10^{-5}$ & $1.8^{+2.1}_{-0.94}\times10^{-5}$
      & $5.0^{+7.5}_{-2.1}\times10^{-5}$ & $1.4^{+1.6}_{-0.73}\times10^{-5}$ \\
    & $85\%$
      & $4.9^{+8.0}_{-2.6}\times10^{-5}$ & $1.5^{+1.8}_{-0.81}\times10^{-5}$
      & $3.6^{+6.3}_{-2.0}\times10^{-5}$ & $1.2^{+1.4}_{-0.63}\times10^{-5}$ \\
    C & $70\%$
      & $7.1^{+8.5}_{-3.5}\times10^{-5}$ & $2.1^{+1.9}_{-0.96}\times10^{-5}$
      & $5.5^{+6.8}_{-2.8}\times10^{-5}$ & $1.6^{+1.5}_{-0.74}\times10^{-5}$ \\
    & $85\%$
      & $5.7^{+5.8}_{-2.7}\times10^{-5}$ & $1.8^{+1.5}_{-0.82}\times10^{-5}$
      & $4.5^{+4.5}_{-2.1}\times10^{-5}$ & $1.4^{+1.2}_{-0.64}\times10^{-5}$ \\
    \addlinespace
    \multicolumn{6}{@{}l}{\emph{plain GNN}} \\
    B & $70\%$
      & $7.4^{+8.0}_{-3.8}\times10^{-4}$ & $2.3^{+2.2}_{-1.1}\times10^{-4}$
      & $6.5^{+6.5}_{-3.3}\times10^{-4}$ & $2.2^{+1.9}_{-1.0}\times10^{-4}$ \\
    & $85\%$
      & $7.0^{+7.7}_{-3.7}\times10^{-4}$ & $2.2^{+2.1}_{-1.1}\times10^{-4}$
      & $6.2^{+6.2}_{-3.1}\times10^{-4}$ & $2.1^{+1.9}_{-0.98}\times10^{-4}$ \\
    C & $70\%$
      & $7.4^{+6.9}_{-3.5}\times10^{-4}$ & $2.4^{+2.0}_{-1.1}\times10^{-4}$
      & $1.4^{+0.99}_{-0.59}\times10^{-3}$ & $5.1^{+3.9}_{-2.3}\times10^{-4}$ \\
    & $85\%$
      & $6.4^{+5.4}_{-2.9}\times10^{-4}$ & $2.1^{+1.8}_{-0.96}\times10^{-4}$
      & $1.3^{+0.84}_{-0.53}\times10^{-3}$ & $5.0^{+3.8}_{-2.2}\times10^{-4}$ \\
    \bottomrule
  \end{tabular}
  \caption{Median expected 95\%~CL upper limits on $\mathrm{BR}(t\to uH)$ and $\mathrm{BR}(t\to cH)$ at integrated luminosities of $300$ and $3000~\mathrm{fb}^{-1}$, combining the associated-production and pair-production signal regions. Results are given for both classifiers and the indicated $b$-tagging efficiencies $\varepsilon_b$, using framework~B (exact Poisson counting) and framework~C (profile-likelihood analysis). The superscript and the subscript give the upward and downward shifts from the median defining the $\pm2\sigma$ expected range under the background-only hypothesis.
  \label{tab:br_lumi}}
\end{table}

\subsection{Interpretation and limitations}
\label{sec:discussion}

The sensitivity gain obtained with the GA-GNN is consistent with a more effective use of radiation patterns associated with the colour-singlet Higgs decay. Although signal and background events can contain the same reconstructed objects, the soft-radiation correlations around the Higgs candidate depend on whether its two jets originate from a colour-singlet decay or are connected to other coloured partons. Organising the hadronic constituents relative to this candidate therefore provides a physically motivated representation of information complementary to the reconstructed event kinematics. For the rare signals considered here, the relevant improvement is the greater signal efficiency retained at very small background acceptance, which explains how moderate changes in the overall ROC AUC can accompany substantially stronger exclusion limits.

The most direct comparison of our results with experimental bounds is with the ATLAS and CMS searches for top FCNCs in the same $H\to b\bar b$ final state with one charged lepton. ATLAS expects $\mathrm{BR}(t\to uH)<8.8\times10^{-4}$ and $\mathrm{BR}(t\to cH)<7.6\times10^{-4}$ from top-quark pair production with $139~\mathrm{fb}^{-1}$~\cite{ATLAS:2023mcc}, while CMS expects $1.1\times10^{-3}$ and $8.6\times10^{-4}$ with $137~\mathrm{fb}^{-1}$~\cite{CMS:2021ncc}. The plain GNN reaches a comparable sensitivity (\cref{tab:limits139}), which provides a useful comparison of the overall sensitivity scale. The GA-GNN improves on these limits by about an order of magnitude and lies below the combinations of several Higgs-boson decay channels, for which ATLAS expects $2.4\times10^{-4}$ and $3.0\times10^{-4}$~\cite{ATLAS:2023ujo} and CMS $2.7\times10^{-4}$ and $3.5\times10^{-4}$~\cite{CMS:2024ubt}. This comparison must, however, be interpreted with care. Our analysis relies on a fast detector simulation and on leading-order background simulations, and is performed at $\sqrt{s}=13.6$~TeV rather than 13~TeV. It also omits the parton-shower and hadronisation modelling uncertainties discussed in section~\ref{sec:profile}, whose effects on the classifier acceptances require a dedicated assessment beyond the uncertainties included here. It nevertheless shows that colour flow, once encoded geometrically, can change the sensitivity of the $H\to b\bar b$ channel substantially.

Comparing the three statistical frameworks provides a complementary check of the sensitivity estimates. At $\mathcal{L}=139~\mathrm{fb}^{-1}$ and $\varepsilon_b=70\%$, the GA-GNN results from frameworks~B and~C use the same GA-GNN working points and yield limits that agree to within 10\%, and pseudo-experiments confirm the asymptotic results of framework~C to within 8\% (section~\ref{sec:profile}). Framework~A is weaker by a factor of about 1.7, because the minimum-background requirement $b\ge b_{\min}$ excludes the highest purity working points where most of the gain of the GA-GNN resides. For the plain GNN, the up-quark limits are comparable across the three frameworks, whereas the charm-quark limit is weaker in framework~C which selects a tighter $t\bar t$ threshold and uses a larger background-normalisation uncertainty. The different background levels at the selected working points also explain the response to changes in the assumed $b$-tagging efficiency. Within the common $\varepsilon_{\rm tag}=\varepsilon_b^3$ rescaling adopted here, raising the per-jet $b$-tagging efficiency from 70\% to 85\% increases $\varepsilon_{\rm tag}$ from $0.34$ to $0.61$, and scales the signal and background yields by the same factor. For the GA-GNN in framework~B, the limits improve by about 30\%, from $1.1\times10^{-4}$ to $7.6\times10^{-5}$ for $t\to uH$ and from $8.3\times10^{-5}$ to $5.8\times10^{-5}$ for $t\to cH$. For the plain GNN in framework~A, they improve by only 15\% and 5\%. At the high-purity working points of the GA-GNN, with backgrounds of one or two events, the limit is driven by the signal yield and benefits directly from a larger tagging efficiency. At the looser working points of the plain GNN, the analysis is closer to the background-dominated and systematics limited regime, in which a common rescaling of signal and background brings little improvement.

The interpretation of the projections to $3000~\mathrm{fb}^{-1}$ also depends on the assumed reduction of the relative uncertainties entering the statistical models. For the experimental uncertainties, this corresponds to calibrations that improve with the data. Since they are the smallest uncertainties of the model, this choice has little impact on the limits. At the working points of the GA-GNN, the background then grows from one or two events to about 25--50 events per signal region, and a 50\% normalisation uncertainty kept at its present size would exceed the statistical uncertainty. The projections therefore assume that the background normalisation can be constrained in situ, for instance from control regions at lower classifier scores, with a precision that improves with the data, and that the simulated samples grow accordingly. At fixed classifier working points, retaining the present relative background uncertainties would eventually limit the improvement obtainable from increasing the luminosity.

Finally, the most sensitive working points of the GA-GNN lie at false positive rates of $10^{-5}$ to $2\times10^{-5}$, where the background estimate rests on one or two simulated events. The finite-sample uncertainty is included through the MC statistical uncertainty and the zero-event prescription of section~\ref{sec:yields}. The working point scan of \cref{fig:br_vs_wp} nevertheless shows that the improvement persists over a broader range of working points: at $\mathrm{FPR}=10^{-4}$ where about ten simulated background events survive, the combined median expected limits amount to $1.8\times10^{-4}$ for $t\to uH$ and $1.4\times10^{-4}$ for $t\to cH$, and remain well below those obtained with the plain-GNN limits at any working points shown.

\section{Conclusion}
\label{sec:conclusion}

In this work, we have shown that a geometric representation of hadronic radiation can help a neural network exploit colour-flow information complementary to the reconstructed event kinematics, substantially improving the sensitivity in the top-quark FCNC benchmark considered here.

The hadronic constituents of each event are mapped from the $(\eta,\phi)$  plane onto the unit sphere using a modified inverse stereographic projection. This mapping is centred on the reconstructed Higgs candidate, with a reference radius set by half the separation of its two assigned $b$-jets in the $(\eta,\phi)$ plane, so that radiation inside and outside the corresponding circular region populates opposite hemispheres. The expected inside-outside asymmetry of soft radiation for a colour singlet  $b\bar b$ pair is therefore promoted to a global geometric feature of the event, which a geodesic-aware GNN exploits using geodesic distances on the sphere to define both the graph neighbourhoods and the edge attributes. To assess the gain in using this representation, we compare the resulting classifier with a baseline GNN defined in the $(\eta,\phi)$ plane, using the same architectures for the kinematic branch and classification head in both cases. As a benchmark, we consider flavour changing neutral interactions of the top quark with the Higgs boson at $\sqrt{s}=13.6$~TeV, in a final state containing one charged lepton, missing transverse momentum and several jets, at least three of which are $b$-tagged. We combine mutually exclusive signal regions targeting associated production, $qg\to tH$, and top-quark pair production with $t\to qH$, where $q=u,c$, using three statistical treatments to assess how the expected exclusion limits depend on small event counts and on the treatment of uncertainties.

For an assumed integrated luminosity of $139~\mathrm{fb}^{-1}$ and a per-jet $b$-tagging efficiency of 70\%, our main profile-likelihood analysis yields median expected 95\%~CL upper limits of $\mathrm{BR}(t\to uH)\simeq1.1\times10^{-4}$ and $\mathrm{BR}(t\to cH)\simeq8\times10^{-5}$ when the associated-production and pair-production  signal regions are combined. These limits improve on the corresponding baseline plain GNN by approximately one order of magnitude or more, and agree within 10\% with those obtained using an exact Poisson treatment. Here, the gain is driven by the high-purity region of the classifier output where the GA-GNN retains five to twelve times as much signal as the plain GNN at a fixed background acceptance of $10^{-4}$, despite more modest changes in the overall ROC AUC. While the baseline GNN achieves a sensitivity comparable to existing related ATLAS and CMS searches in the $H\to b\bar b$ channel, the GA-GNN projections are stronger than the expected Run-2 limits stemming from combinations of several Higgs-boson decay channels. Moreover, increasing the $b$-tagging efficiency from 70\% to 85\% strengthens the limits at $139~\mathrm{fb}^{-1}$ by about 30\%. Finally, for $\varepsilon_b=70\%$, the projected limits reach $\mathrm{BR}(t\to uH)<2.1\times10^{-5}$ and $\mathrm{BR}(t\to cH)<1.6\times10^{-5}$ at $3000~\mathrm{fb}^{-1}$, provided that background constraints and experimental calibrations improve and the simulated sample sizes increase as assumed.

These limits illustrate the potential of the method and should not be over-interpreted, as several effects are treated in a simplified manner. The response of the detector is simulated using a fast parametrisation, background events are generated at leading order, with a common uncertainty assigned to the total background normalisation, the classifier acceptance is evaluated for the combined background sample, and the most sensitive working points contain only  a small number of simulated background events. Most importantly, uncertainties  associated with the modelling of the parton shower and hadronisation, to which  a classifier exploiting colour flow is directly sensitive, have not been included. A priority for further work is therefore a quantitative assessment of how parton-shower and hadronisation modelling affects the classifier acceptance, following approaches developed for particle level  spectra~\cite{Amoroso:2018qga, Jueid:2022qjg, Jueid:2023vrb} and  constrained by colour flow measurements~\cite{D0:2011lzz, ATLAS:2015ytt, ATLAS:2018olo}. Larger background samples would also improve the stability of the high purity region, while storing the classifier output separately for each process would allow the corresponding background uncertainties to be assigned independently. A complete assessment at the high luminosity LHC~\cite{Cepeda:2019klc} will therefore require both a robust treatment of these modelling uncertainties and in-situ constraints on the backgrounds.

More broadly, the same representation can be applied to any colour singlet decay into a quark-antiquark pair, including $H\to b\bar b$ and $H\to c\bar c$ in associated production, Higgs-boson pair production, and FCNC top-quark interactions involving a hadronically decaying $Z$ boson. It can also be could be adapted to other colour-singlet decays into two gluons. These results motivate the use of geometric representations of QCD radiation alongside reconstructed event kinematics to improve sensitivity to rare processes with hadronic final states.

\section*{Acknowledgements}
The work of A.~H.~and A.~J.~is supported by Korea Institute for Advanced
Study (KIAS) grant funded by the Korean government. A.~J.~would like to thank the CERN Theoretical Physics Department for hospitality during the completion of this work. M.~N.~is funded by 22H05113, “Foundation of Machine Learning Physics”, Grant-in-Aid for Transformative Research Areas, and  JPMJAR2633 JST ARiSE(AI4MP). A.~H.~is partially supported by the Science, Technology \& Innovation Funding Authority (STDF) under project ID 50806.

\bibliography{bibliography}

\nolinenumbers

\end{document}